\documentclass[a4paper,onecolumn,11pt]{quantumarticle}
\pdfoutput=1
\usepackage{graphicx}
\usepackage{amsmath}
\usepackage{amsthm}
\usepackage{amssymb}
\usepackage{latexsym}
\usepackage{array}
\usepackage{hyperref}
\usepackage{float}
\usepackage{amsfonts}
\usepackage{xcolor}

\newcommand{\bra}[1]{\left<#1\right|}
\newcommand{\ket}[1]{\left|#1\right>}

\newcommand{\probP}{\text{I\kern-0.15em P}}

\newcommand{\BE}{\begin{equation}}
\newcommand{\EE}{\end{equation}}
\newcommand{\BEA}{\begin{eqnarray}}
\newcommand{\EEA}{\end{eqnarray}}

\newtheorem{proposition}{Proposition}

\newtheorem{corollary}{Corollary}

\begin{document}

\title{Finite Imaginary-Time Evolution for Polynomial Unconstrained Binary Optimization}

\author{Jaehee Kim}
\affiliation{Sungkyunkwan University Advanced Institute of Nanotechnology, Suwon, Korea}

\author{Juhyeon Kim}
\affiliation{Department of Semiconductor Convergence Engineering, Sungkyunkwan University, Suwon, Korea}

\author{Gwonhak Lee}
\affiliation{IBM Quantum, Seoul 07335, Republic of Korea}

\author{Kyunghyun Baek}
\affiliation{Institute for Convergence Research and Education in Advanced Technology, Yonsei University, Seoul 03722, Republic of Korea}
\affiliation{Department of Quantum Information, Yonsei University, Seoul, 03722, Republic of Korea}

\author{Daniel K. Park}
\affiliation{Department of Applied Statistics, Yonsei University, Seoul, 03722, Republic of Korea}
\affiliation{Department of Statistics and Data Science, Yonsei University, Seoul, 03722, Republic of Korea}
\affiliation{Department of Quantum Information, Yonsei University, Seoul, 03722, Republic of Korea}

\author{Jeongho Bang}\email{jbang@yonsei.ac.kr}
\affiliation{Institute for Convergence Research and Education in Advanced Technology, Yonsei University, Seoul 03722, Republic of Korea}
\affiliation{Department of Quantum Information, Yonsei University, Seoul, 03722, Republic of Korea}

\author{Joonsuk Huh}\email{joonsukhuh@yonsei.ac.kr}
\affiliation{Department of Chemistry, Yonsei University, Seoul 03722, Republic of Korea}
\affiliation{Department of Quantum Information, Yonsei University, Seoul, 03722, Republic of Korea}

\begin{abstract}
Imaginary-time evolution is a standard primitive for ground-state preparation but is nonunitary, precluding direct quantum implementation. We develop Finite Imaginary-Time Evolution (FinITE), a finite-$\beta$ construction for diagonal Pauli-$Z$ cost Hamiltonians arising from polynomial unconstrained binary optimization (PUBO) instances, including QUBO and HUBO cases. FinITE uses the linear-combination-of-unitaries (LCU) framework to implement a scaled imaginary-time propagator. The commuting Pauli-$Z$ structure makes termwise block-encodings compose without product-formula error, and higher-order Pauli-$Z$ terms are handled directly without quadratization. The structure yields an exact finite-$\beta$ identity between the LCU success probability and the ground-subspace fidelity. Combined with a gap-based fidelity lower bound, the identity yields a closed-form sufficient imaginary-time threshold $\beta^\star$ for a chosen target fidelity. The threshold depends on estimates of the spectral gap and the initial ground-subspace overlap. Because the LCU success event is flagged by a known ancilla outcome, we integrate fixed-point amplitude amplification with an explicit query-complexity bound. Statevector simulations verify the identity on a five-vertex MaxCut (QUBO) and an eight-qubit cubic HUBO instance, and shot-based simulations on the MaxCut instance illustrate the predicted finite-$\beta$ threshold and amplification procedure.
\end{abstract}

\maketitle

%===============================================================
\section{Introduction}
\label{sec:intro}
%===============================================================

Combinatorial optimization underlies decision problems across logistics, resource allocation, scheduling, engineering design, and data-driven model selection~\cite{kochenberger2014unconstrained,abbas2024challenges}. A broad and widely studied class of such problems is polynomial unconstrained binary optimization (PUBO), in which a polynomial objective function of binary variables is minimized over the Boolean hypercube. PUBO admits a standard mapping to a diagonal cost Hamiltonian whose ground state encodes the optimum~\cite{lucas2014ising}. For this reason, PUBO has been a central target for quantum optimization algorithms. Objectives may be quadratic, as in quadratic unconstrained binary optimization (QUBO), or have maximum polynomial degree at least three, as in higher-order unconstrained binary optimization (HUBO). Reducing HUBO to QUBO through quadratization introduces auxiliary variables that enlarge the search space and alter the cost structure.

Imaginary-time evolution (ITE) is a standard primitive for ground-state preparation. The operator $e^{-\beta\hat H}$ exponentially suppresses excited-state components. Given any initial state with nonzero ground-subspace overlap, the normalized output concentrates on the ground subspace. This makes imaginary-time evolution useful for optimization. Classical routes to simulating this operator include sampling-based methods such as quantum Monte Carlo~\cite{foulkes2001qmc} and variational-compression methods such as tensor-network and density-matrix renormalization-group techniques~\cite{vidal2004efficient,schollwoeck2011dmrg}. The key difficulty is that $e^{-\beta\hat H}$ is nonunitary, so it cannot be implemented directly by a unitary quantum circuit.

Several algorithms have been developed to realize imaginary-time evolution on quantum hardware. Variational imaginary-time evolution (VITE) projects the imaginary-time dynamics onto a parameterized circuit manifold, with parameter updates derived from a variational principle on a classical computer~\cite{mcardle2019variational,bauer2023combinatorial}. Quantum imaginary-time evolution (QITE) approximates each imaginary-time step by a local unitary reconstruction within correlation-bounded domains~\cite{motta2020determining}. Probabilistic imaginary-time evolution (PITE) exploits ancilla measurements on an expanded Hilbert space to realize the non-unitary operator~\cite{kosugi2022imaginarytime,xie2024probabilistic}. Recent block-encoded LCU variants for Pauli-product Hamiltonians include Rrapaj and Rule (universal quantum neural networks)~\cite{rrapaj2024exact}, Zhong \textit{et al.} (MaxCut with QAOA hybridization)~\cite{zhong2024classical}, and Yi \textit{et al.} (single-ancilla Taylor-LCU construction)~\cite{yi2025probabilistic}.

These recent constructions operate at finite $\beta$ and provide post-selected implementations. Within this finite-$\beta$ regime, the remaining algorithmic task is to choose $\beta$ for a target ground-subspace fidelity while simultaneously tracking the LCU success probability induced by the chosen block-encoding subnormalization. The prior constructions do not develop, in closed form, the joint relation between this success probability and the ground-subspace fidelity, the resulting $\beta$-selection rule, or the fixed-point amplitude-amplification query cost at the selected $\beta$. FinITE provides this closed-form analysis for diagonal Pauli-$Z$ PUBO Hamiltonians.

In this work we develop Finite Imaginary-Time Evolution (FinITE), a finite-$\beta$ LCU-based construction for diagonal Pauli-$Z$ cost Hamiltonians arising from PUBO instances. FinITE uses a termwise LCU block-encoding~\cite{childs2012Hamiltonian} of a scaled $e^{-\beta\hat H}$, in which each Pauli-$Z$ string contributes an analytically determined block. Because the strings commute, the termwise blocks compose without product-formula error. Higher-order Pauli-$Z$ terms are handled directly, without quadratization. We then prove the exact finite-$\beta$ identity $P_{\mathrm{LCU}}(\beta)\,F_g(\beta) = \gamma_0\,e^{-2\beta(W+E_0)}$. Here $P_{\mathrm{LCU}}$ and $F_g$ are the LCU success probability and the ground-subspace fidelity, respectively. The remaining symbols denote the initial ground-subspace overlap $\gamma_0$, the $\ell_1$-norm $W$ of the nonzero Pauli coefficients, and the ground-state energy $E_0$. The identity quantifies the trade-off between the LCU success probability and the ground-subspace fidelity at any finite $\beta$. Combined with a gap-based fidelity lower bound, the identity yields a closed-form sufficient imaginary time $\beta^\star(\bar F)$ for a chosen target fidelity $\bar F$. The threshold depends on estimates of the spectral gap $\Delta$ and the initial overlap $\gamma_0$. At $\beta^\star$ the LCU success probability can be the resource bottleneck. Because the LCU success event is flagged by a known ancilla outcome, we apply fixed-point amplitude amplification~\cite{grover2005fixed,yoder2014fixed} to that event and obtain an explicit query-complexity bound at $\beta^\star$. We verify the identity to near machine precision on a five-vertex MaxCut (QUBO) and an eight-qubit cubic Pauli-$Z$ HUBO instance. Shot-based simulations illustrate the predicted finite-$\beta$ fidelity threshold and the amplification procedure on the MaxCut instance.

The paper is organized as follows. Sec.~\ref{sec:background} introduces the PUBO Hamiltonian model, including its QUBO and HUBO special cases, and the LCU and FPAA primitives used later. Sec.~\ref{sec:construction} develops the termwise-LCU construction and analyzes its large-$\beta$ limit. Sec.~\ref{FinITE} presents FinITE, derives the finite-$\beta$ identity and the threshold rule, and describes the FPAA integration. Sec.~\ref{POPND} reports the numerical verification. Sec.~\ref{DISC} concludes with limitations and future directions.
%%%%%%%%%%%%%%%%%%%%%%%%

\section{Background}\label{sec:background}

This section fixes notation and briefly reviews the standard LCU and FPAA ingredients used later; readers already familiar with these preliminaries may proceed to Sec.~\ref{sec:construction}. A consolidated symbol list is provided in Appendix~\ref{app:notation}.

\subsection{Polynomial Unconstrained Binary Optimization}

A PUBO instance is specified by a polynomial cost function over binary variables $x_i \in \{0,1\}$. Quadratic unconstrained binary optimization (QUBO) denotes PUBO instances of polynomial degree at most two and minimizes
\begin{equation}
C(x) \;=\; \sum_{i=1}^n h_i\, x_i + \sum_{i<j} J_{ij}\, x_i x_j \;=\; x^T Q x,
\end{equation}
where the right-hand side absorbs the linear coefficients $h_i$ into the diagonal of the symmetric matrix $Q$. Higher-order unconstrained binary optimization (HUBO) denotes PUBO instances whose maximum polynomial degree is at least three,
\begin{equation}
C(x) \;=\; \sum_i h_i x_i + \sum_{i<j} J_{ij} x_i x_j + \sum_{i<j<k} K_{ijk}\, x_i x_j x_k + \ldots.
\end{equation}
Substituting $x_i = (I - \sigma_z^{(i)})/2$ turns the cost function into a Hamiltonian whose ground state encodes the optimum up to an additive identity shift:
\begin{equation}
\hat H_{\mathrm{QUBO}} = \sum_i h_i\, \frac{I - \sigma_z^{(i)}}{2} + \sum_{i<j} J_{ij}\, \frac{I - \sigma_z^{(i)}}{2}\frac{I - \sigma_z^{(j)}}{2},
\end{equation}
\begin{equation}
\hat H_{\mathrm{HUBO}} = \sum_{\alpha} c_{\alpha} \prod_{i \in S_\alpha} \frac{I - \sigma_z^{(i)}}{2}.
\end{equation}
Throughout we use the $0/1$ encoding above rather than a $\pm 1$ Ising convention. Expanding the $(I - \sigma_z)/2$ factors yields a Pauli-$Z$ decomposition $\hat H = \sum_{\boldsymbol\mu} x_{\boldsymbol\mu} \sigma_{\boldsymbol\mu}$. For $\boldsymbol\mu = 0$, $\sigma_{\boldsymbol\mu} = I$, so this term is a scalar shift $c\,I$. We remove this identity term from the LCU and account for it classically by multiplying the final state by $e^{-\beta c}$. This scalar cancels in the normalized post-selected fidelity. Thus the $\ell_1$-norm $W$ used below is $W = \sum_{\boldsymbol\mu \ne 0} |x_{\boldsymbol\mu}|$ and excludes the identity component. Both $\hat H_{\mathrm{QUBO}}$ and $\hat H_{\mathrm{HUBO}}$ are diagonal in the computational basis, $\hat H_{\mathrm{PUBO}} = \sum_{x \in \{0,1\}^n} E_x |x\rangle\langle x|$, and all Pauli-$Z$ strings in their expansions pairwise commute.

HUBO can in principle be reduced to QUBO by introducing auxiliary variables. For a cubic term $x_i x_j x_k$, Rosenberg's gadget~\cite{rosenberg1975reduction} introduces an auxiliary variable $y$ and enforces $y = x_j x_k$ via the penalty $P\,(x_j x_k - 2 x_j y - 2 x_k y + 3 y)$, which vanishes if and only if $y = x_j x_k$ on binary inputs. Applying the gadget iteratively to a $k$-local HUBO on $n$ variables costs $\mathcal O(k)$ auxiliary variables for each interaction. In the worst case of a dense order-$k$ cost function, this gives $\mathcal O(n^k)$ auxiliary variables in total and generically produces a denser problem structure. Our LCU construction (Sec.~\ref{imp_qc}) block-encodes the higher-order terms directly, so we do not apply this quadratization step.

Real-time evolution under the cost Hamiltonian $\hat H$ does not change populations in its energy eigenbasis, and in particular preserves the overlap with the ground eigenspace. Cost-Hamiltonian evolution alone therefore cannot increase the probability of measuring an optimal configuration. In this paper, the mechanism beyond such unitary evolution is the imaginary-time operator $e^{-\beta\hat H}$, which nonunitarily reweights energy components toward lower energies. We implement it with the LCU framework introduced below and analyze it in Sec.~\ref{imp_qc}, with the large-$\beta$ breakdown treated in Sec.~\ref{Issue_with_ITE}.

\subsection{Linear Combination of Unitaries}
\label{subsec:LCU_primer}

The imaginary-time operator $e^{-\beta\hat H}$ is non-unitary and cannot be applied directly as a quantum circuit. The Linear Combination of Unitaries (LCU) framework~\cite{childs2012Hamiltonian} realizes a weighted sum such as $\kappa U_a + U_b$ with $\kappa \ge 0$. It uses an ancilla qubit, controlled unitary selection, and post-selection on the ancilla outcome $|0\rangle$. Conditioned on this outcome, the implemented operator is
\[
\frac{\kappa U_a + U_b}{\kappa+1},
\]
so the target operator is obtained with subnormalization factor $\kappa+1$. In block-encoding language~\cite{Gilyen2019quantum}, post-selection realizes a scaled block of the target non-unitary operator, with subnormalization determined by the LCU coefficient sum. The explicit circuit and success-probability expression are given in Appendix~\ref{app:LCU}. The application to our commuting PUBO Hamiltonian is developed in Sec.~\ref{imp_qc}. The block-encoding perspective is revisited in Sec.~\ref{Issue_with_ITE}.

\subsection{Fixed-Point Amplitude Amplification}
\label{subsec:FPAA_primer}

Standard amplitude amplification~\cite{brassard2002quantum} achieves a quadratic speedup, but fixed-iteration schedules typically require an estimate of the initial target amplitude to avoid overshooting the optimal rotation angle. In LCU-based imaginary-time evolution the target subspace is known: success corresponds to measuring the ancilla register in the all-zero state. The corresponding initial target amplitude, however, is generically unknown in advance. Fixed-Point Amplitude Amplification (FPAA)~\cite{yoder2014fixed} avoids this overshooting via a fixed-point schedule of phased reflections. Given any positive lower bound $\lambda > 0$ on the initial squared target amplitude, FPAA boosts the success probability to at least $1 - \delta^2$ using
\begin{equation}
L \;=\; \mathcal O\!\left(\frac{\log(2/\delta)}{\sqrt{\lambda}}\right)
\label{eq:FPAA_query}
\end{equation}
applications of the amplification primitives. The explicit phase schedule and convergence proof are given in~\cite{yoder2014fixed}. Appendix~\ref{app:FPAA} gives the circuit realization used in this work. The FinITE specialization and the resulting resource counts are developed in Sec.~\ref{FinITE} and Appendix~\ref{app:cnot_complexity}.

\section{Termwise-LCU Construction and Large-$\beta$ Limit}
\label{sec:construction}
\label{imp_qc}

Sec.~\ref{sec:background} introduced the PUBO Hamiltonian $\hat H = \sum_{\boldsymbol\mu} x_{\boldsymbol\mu} \sigma_{\boldsymbol\mu}$, whose Pauli terms pairwise commute. It also introduced the LCU and FPAA primitives used in the following sections. This section combines these ingredients to construct a block-encoding of the imaginary-time operator $e^{-\beta\hat H}$, up to a scalar factor. Each commuting Pauli term is normalized into its own LCU block. The probability of the joint post-selection event is denoted $P_{\mathrm{LCU}}(\beta)$. Sec.~\ref{FinITE} then combines this construction with FPAA to form the full FinITE algorithm and relates $P_{\mathrm{LCU}}(\beta)$ exactly to the ground-subspace fidelity $F_g(\beta)$.

\subsection{Rationale for termwise LCU}
\label{subsec:approach}

For commuting PUBO Hamiltonians, FinITE uses the LCU block-encoding route of Sec.~\ref{subsec:LCU_primer}. This choice gives two advantages for the imaginary-time operator $e^{-\beta\hat H}$. (i) It yields an exact, non-variational implementation of the scaled finite-$\beta$ ITE operator with explicit resource bounds. This avoids the local-unitary fitting and approximation errors inherent in QITE-style constructions. (ii) The circuit structure of each LCU block-encoding is independent of $\beta$. Only the rotation angles depend on $\beta$, so the same compiled circuit can be reused across the $\beta$ range with classical parameter updates. The overall FinITE resource cost nevertheless grows with $\beta$. A larger $\beta$ lowers the LCU success probability $P_{\mathrm{LCU}}(\beta)$, which inflates the required FPAA query complexity (Sec.~\ref{FinITE}, Appendix~\ref{app:cnot_complexity}).

\subsection{Commuting ITE factorization}
\label{subsec:factorization}

Because the Pauli terms in $\hat H$ commute, the ITE operator factorizes into termwise exponentials that are block-encoded by the termwise LCU of Sec.~\ref{subsec:termwise_LCU_sub}. The normalized imaginary-time state is
\begin{align}
\label{eq:ITE_def}
    \ket{\psi(\beta)} = \frac{e^{-\beta\hat H}\ket{\psi_0}}{\sqrt{\bra{\psi_0}e^{-2\beta\hat H}\ket{\psi_0}}},
\end{align}
As $\beta \to \infty$, this state converges to $\Pi_{\mathcal G}\ket{\psi_0}/\sqrt{\gamma_0}$, where $\Pi_{\mathcal G}$ is the orthogonal projector onto the ground subspace $\mathcal G$. The prefactor $\gamma_0 := \bra{\psi_0}\Pi_{\mathcal G}\ket{\psi_0}$ is the initial overlap with the ground subspace, assumed nonzero. Appendix~\ref{app:ITE} derives this result by Wick rotation and eigenbasis expansion.

For the commuting PUBO Hamiltonian $\hat H = \sum_{\boldsymbol\mu} x_{\boldsymbol\mu}\sigma_{\boldsymbol\mu}$ of Sec.~\ref{sec:background}, the ITE operator factorizes exactly over the nonzero-coefficient terms. Each factor expands via $\sigma_{\boldsymbol\mu}^2 = I$:
\begin{align}
\label{exact_trotterized_more_expand}
\begin{split}
    \exp(-\beta\hat H) &= \prod_{\boldsymbol\mu} \exp(-\beta x_{\boldsymbol\mu}\sigma_{\boldsymbol\mu}) \\
    &= \prod_{\boldsymbol\mu} \Bigl[\cosh(\beta|x_{\boldsymbol\mu}|)\,I - \mathrm{sgn}(x_{\boldsymbol\mu})\sinh(\beta|x_{\boldsymbol\mu}|)\,\sigma_{\boldsymbol\mu}\Bigr].
\end{split}
\end{align}
The factorization is exact because the Pauli strings pairwise commute; no Trotter error enters. The construction uses one LCU block for each nonzero Pauli term, for a total of $M$ blocks.

\subsection{Normalized termwise-LCU construction}
\label{subsec:termwise_LCU_sub}

Block-encoding via LCU requires the encoded operator to have operator norm at most $1$ (Sec.~\ref{subsec:LCU_primer}). In a termwise LCU, each Pauli factor $e^{-\beta x_{\boldsymbol\mu}\sigma_{\boldsymbol\mu}}$ is block-encoded separately, and each has op-norm $e^{\beta|x_{\boldsymbol\mu}|} > 1$ for nonzero $x_{\boldsymbol\mu}$. The subnormalizations of the individual terms therefore determine an overall scaling $m(\beta)$, and the joint block-encoding realizes $m(\beta)\,e^{-\beta\hat H}$. Post-selection on the LCU ancilla then realizes the normalized ITE state~\eqref{eq:ITE_def}, bypassing the classical evaluation of $\sqrt{\bra{\psi_0}e^{-2\beta\hat H}\ket{\psi_0}}$.

We scale each factor in the product~\eqref{exact_trotterized_more_expand} so that it matches the two-term LCU primitive $(\kappa U_a + U_b)/(\kappa+1)$ of Sec.~\ref{subsec:LCU_primer} with nonnegative coefficients. Restricted to nonzero-coefficient terms, this gives
\[
m(\beta) = \prod_{\boldsymbol\mu:\,x_{\boldsymbol\mu}\ne 0} e^{-\beta|x_{\boldsymbol\mu}|} = e^{-\beta W}.
\]

Concretely, for $\beta > 0$ and each nonzero-coefficient term $\boldsymbol\mu$, set
\[
\kappa_{\boldsymbol\mu} = \coth(\beta|x_{\boldsymbol\mu}|), \qquad U_a = I, \qquad U_b = -\mathrm{sgn}(x_{\boldsymbol\mu})\,\sigma_{\boldsymbol\mu}.
\]
The LCU primitive then gives
\[
\frac{\kappa_{\boldsymbol\mu}\,I - \mathrm{sgn}(x_{\boldsymbol\mu})\sigma_{\boldsymbol\mu}}{\kappa_{\boldsymbol\mu}+1}
\;=\; \frac{\cosh(\beta|x_{\boldsymbol\mu}|)\,I - \mathrm{sgn}(x_{\boldsymbol\mu})\sinh(\beta|x_{\boldsymbol\mu}|)\,\sigma_{\boldsymbol\mu}}{e^{\beta|x_{\boldsymbol\mu}|}}.
\]
The denominator $\kappa_{\boldsymbol\mu}+1 = (\cosh(\beta|x_{\boldsymbol\mu}|) + \sinh(\beta|x_{\boldsymbol\mu}|))/\sinh(\beta|x_{\boldsymbol\mu}|)$ is the LCU subnormalization for this two-term block. The per-term block therefore contributes a denominator $e^{\beta|x_{\boldsymbol\mu}|}$ to the scaled operator, so the total subnormalization across all $M$ blocks is $\alpha_{\mathrm{total}}(\beta) = \prod_{\boldsymbol\mu \ne 0} e^{\beta|x_{\boldsymbol\mu}|} = e^{\beta W}$, the $\ell_1$-norm factor. Taking the product over nonzero-coefficient terms yields
\begin{align}
\label{eq:termwise_LCU}
    m(\beta)\,\exp(-\beta\hat H) = \prod_{\boldsymbol\mu:\,x_{\boldsymbol\mu}\ne 0} \frac{\kappa_{\boldsymbol\mu}\,I - \mathrm{sgn}(x_{\boldsymbol\mu})\sigma_{\boldsymbol\mu}}{\kappa_{\boldsymbol\mu}+1}.
\end{align}
The construction succeeds when every ancilla is measured in $|0\rangle$. The probability of this joint event, $P_{\mathrm{LCU}}(\beta)$, is computed in Proposition~\ref{thm:exact_identity} of Sec.~\ref{FinITE} together with its relation to the ground-subspace fidelity $F_g(\beta)$.

\subsection{Large-$\beta$ limit and its breakdown}
\label{subsec:collapse}
\label{Issue_with_ITE}

The termwise-LCU construction~\eqref{eq:termwise_LCU} has a simple large-$\beta$ limit. Each block converges to the orthogonal projector $(I - \mathrm{sgn}(x_{\boldsymbol\mu})\sigma_{\boldsymbol\mu})/2$ onto the $\sigma_{\boldsymbol\mu}$-eigenspace with eigenvalue $-\mathrm{sgn}(x_{\boldsymbol\mu})$, and
\begin{align}
\label{ITE_infty}
    \lim_{\beta\to\infty} m(\beta)\,\exp(-\beta\hat H) = \prod_{\boldsymbol\mu} \frac{I - \mathrm{sgn}(x_{\boldsymbol\mu})\sigma_{\boldsymbol\mu}}{2}.
\end{align}
Although each factor has rank $2^{n-1}$, their product projects onto the simultaneous eigenspace satisfying all sign constraints. In the common eigenbasis $\{|\lambda^{(k)}\rangle\}$,
\begin{align}
\label{ITE_infty_eigenbasis}
    \prod_{\boldsymbol\mu}\frac{I-\mathrm{sgn}(x_{\boldsymbol\mu})\sigma_{\boldsymbol\mu}}{2} = \sum_k \prod_{\boldsymbol\mu} \frac{1-\mathrm{sgn}(x_{\boldsymbol\mu})E_{\boldsymbol\mu}^{(k)}}{2}\, |\lambda^{(k)}\rangle\langle\lambda^{(k)}|.
\end{align}
Thus any eigenstate with $E_{\boldsymbol\mu}^{(k)} = \mathrm{sgn}(x_{\boldsymbol\mu})$ for even one term is annihilated; the limiting operator is nonzero only on common eigenstates where every $\sigma_{\boldsymbol\mu}$ has eigenvalue $-\mathrm{sgn}(x_{\boldsymbol\mu})$.

This limit is operationally useless in general. For any exact block-encoding of $e^{-\beta\hat H}$ with subnormalization $\alpha(\beta) \ge \|e^{-\beta\hat H}\|_{\mathrm{op}} = e^{-\beta E_0}$, the success probability $P_{\mathrm{succ}}(\beta)$ and the ground-subspace fidelity $F_g(\beta)$ of the normalized post-selected state satisfy
\begin{equation}
P_{\mathrm{succ}}(\beta) \cdot F_g(\beta) \;=\; \frac{\gamma_0\, e^{-2\beta E_0}}{\alpha(\beta)^2}.
\label{eq:BE_identity_general}
\end{equation}
Indeed, exact block-encoding gives $P_{\mathrm{succ}}(\beta) = \|e^{-\beta\hat H}|\psi_0\rangle\|^2/\alpha(\beta)^2$, while the eigenbasis expansion of the normalized post-selected state gives $F_g(\beta) = \gamma_0\, e^{-2\beta E_0}/\|e^{-\beta\hat H}|\psi_0\rangle\|^2$; multiplying cancels the shared factor $\|e^{-\beta\hat H}|\psi_0\rangle\|^2$.
Thus the $\beta\to\infty$ limit is useful only if $\alpha(\beta)$ tracks the optimal scale $e^{-\beta E_0}$ up to a bounded factor. Choosing such a normalization would require prior knowledge of $E_0$, which is the very quantity the algorithm seeks.

FinITE uses the subnormalization $\alpha_{\mathrm{LCU}}(\beta) = e^{\beta W}$ fixed by the termwise-LCU construction of Sec.~\ref{subsec:termwise_LCU_sub}. This value comes from the universal triangle-inequality bound $\|\hat H\|_{\mathrm{op}} \le W$ (equivalently, $E_0 \ge -W$), not from the true ground-state energy, which is unknown. As $\beta \to \infty$, the normalized operator approaches the projector of Eq.~\eqref{ITE_infty}, whose range is the simultaneous eigenspace where every $\sigma_{\boldsymbol\mu}$ has eigenvalue $-\mathrm{sgn}(x_{\boldsymbol\mu})$. This eigenspace coincides with the ground space only when the bound is saturated ($E_0 = -W$). For generic problems $E_0 > -W$, the simultaneous eigenspace is typically empty or exponentially small, and the LCU success probability collapses.

\paragraph*{MAX-CUT illustration.}
For MAX-CUT with $\hat H = \sum_{(i,j)\in E} \sigma_z^{(i)}\sigma_z^{(j)}$~\cite{hadfield2021on,farhi2014quantum,wecker2016training}, Eq.~\eqref{ITE_infty} specializes to
\begin{equation}
\label{Max-Cut-op}
    \prod_{(i,j)\in E} \frac{I - \sigma_z^{(i)}\sigma_z^{(j)}}{2},
\end{equation}
which is non-zero only on basis states cutting every edge; such labelings exist if and only if the graph is bipartite. For the uniform initial state $\ket{\psi_0} = \ket{+}^{\otimes n}$, $P_{\mathrm{LCU}}(\beta\to\infty) = N_{\mathrm{cut}}/2^n$ where $N_{\mathrm{cut}}$ is the number of such labelings. Non-bipartite graphs give $N_{\mathrm{cut}}=0$, so the limiting probability vanishes. Connected bipartite graphs give $N_{\mathrm{cut}}=2$, so $P_{\mathrm{LCU}}(\beta\to\infty) = 2^{1-n}$, still exponentially small. Thus the $\beta\to\infty$ limit gives no useful regime, and a finite $\beta$ must be chosen.

The operator at finite $\beta$ does not vanish, but its LCU success probability is traded off against the conditional ground-subspace fidelity. The exact form of that trade-off, for any initial state and any finite $\beta$, is Proposition~\ref{thm:exact_identity} in Sec.~\ref{FinITE}.

\section{Finite Imaginary-Time Evolution}\label{FinITE}
%-------------------------------------------------------- ---------------------------------------------------------
As established in Sec.~\ref{Issue_with_ITE}, the termwise-normalized LCU implementation of ITE collapses as $\beta \to \infty$ on commuting Hamiltonians, making the $\beta \to \infty$ limit useless in practice. FinITE replaces this limit with a finite, analytically chosen $\beta$ and compensates the reduced LCU success probability with fixed-point amplitude amplification.

The algorithm has two stages at a fixed $\beta$. \emph{Stage 1} is a block-encoded LCU that applies $\hat M(\beta) = m(\beta)\,e^{-\beta \hat H}$ to the initial state, where $m(\beta) = e^{-\beta W}$ is the scalar normalization fixed by the termwise-LCU construction (see Sec.~\ref{imp_qc}). Success is indicated by all ancillae measuring in $|0\rangle$; the resulting probability $P_{\mathrm{LCU}}(\beta)$ and the ground-subspace fidelity $F_g(\beta)$ of the normalized post-selected state are the two figures of merit.

\emph{Stage 2} boosts $P_{\mathrm{LCU}}(\beta)$ via FPAA~\cite{yoder2014fixed}. If $\lambda$ is any known lower bound on $P_{\mathrm{LCU}}(\beta)$, then $L = \mathcal O(\log(2/\delta)/\sqrt{\lambda})$ iterations suffice to reach the target subspace up to error $\delta$ (phase schedule:~\cite{yoder2014fixed}; circuit conventions: Sec.~\ref{subsec:FPAA_primer} and Appendix~\ref{app:FPAA}). The marked subspace is the LCU-success event, not the optimization solution. FPAA therefore requires no estimate of the ground-subspace overlap---only a lower bound on $P_{\mathrm{LCU}}(\beta)$. The analysis below supplies that lower bound and links $P_{\mathrm{LCU}}(\beta)$ to $F_g(\beta)$.

In resource terms, FinITE uses $n$ system qubits and $M$ termwise-LCU ancillae, one for each Pauli term (Sec.~\ref{imp_qc}); the FPAA good-subspace reflection marks the all-zero ancilla subspace and requires no separate flag register. Appendix~\ref{app:cnot_complexity} uses a multi-controlled phase-kickback decomposition with one additional scratch ancilla, for a total of $n+M+1$ qubits.

Two questions then determine the algorithm's cost: \emph{(a)} how does $P_{\mathrm{LCU}}(\beta)$ depend on $\beta$, and \emph{(b)} how does $F_g(\beta)$ improve with $\beta$? Proposition~\ref{thm:exact_identity} below gives the exact identity between the two for any pairwise-commuting PUBO Hamiltonian and any initial state. It turns the informal bias/feasibility picture into a closed-form rule for picking $\beta$.

\subsection{Exact identity and $\beta$-threshold rule}
\label{subsec:exact_identity}

For PUBO Hamiltonians with pairwise-commuting Pauli terms---the diagonal PUBO case treated in this paper, illustrated numerically on QUBO and HUBO instances---the LCU success probability and the conditional ground-subspace fidelity obey an \emph{exact spectral relation} in terms of the initial-state spectrum. This result can be tighter than the state-independent termwise LCU bound~\cite{childs2012Hamiltonian}. More importantly, it yields a principled rule for choosing $\beta$ given a target fidelity. We state the result here and defer the proof to Appendix~\ref{app:exact_identity}.

Fix a Pauli decomposition $\hat H = \sum_{\boldsymbol\mu} x_{\boldsymbol\mu} \sigma_{\boldsymbol\mu}$ of an $n$-qubit Hamiltonian with real coefficients $x_{\boldsymbol\mu} \in \mathbb R$ and pairwise-commuting Pauli strings $\sigma_{\boldsymbol\mu}$. Following Sec.~\ref{sec:background}, the identity term has been removed and absorbed into the classical scalar $e^{-\beta c}$; throughout this section and its proof (Appendix~\ref{app:exact_identity}) the index $\boldsymbol\mu$ therefore ranges over nonzero-coefficient Pauli strings only, and $\hat H$ denotes the identity-removed part. Let $\{|E_k\rangle\}$ be an orthonormal common eigenbasis with eigenvalues $E_0 \le E_1 \le \cdots \le E_{d-1}$. Let $\mathcal G = \{k : E_k = E_0\}$ denote the (possibly degenerate) set of ground-state indices, and write the normalized initial state as $|\psi_0\rangle = \sum_k c_k |E_k\rangle$ with $\sum_k |c_k|^2 = 1$. Define the cumulative LCU normalization and the initial-state overlap with the ground subspace as
\begin{align}
W \; := \; \sum_{\boldsymbol\mu \ne 0} |x_{\boldsymbol\mu}|, \qquad
\gamma_0 \; := \; \sum_{k \in \mathcal G} |c_k|^2,
\end{align}
so that $m(\beta) = e^{-\beta W}$ is the product of the normalization factors of the termwise-LCU blocks (Sec.~\ref{subsec:termwise_LCU_sub}). Let $P_{\mathrm{LCU}}(\beta)$ denote the probability that all ancillae of the termwise-LCU post-select to $|0\rangle$, and let $F_g(\beta)$ denote the ground-subspace fidelity of the normalized post-selected state.

Before stating the identity, we relate it to the generic block-encoding scaling in Eq.~\eqref{eq:BE_identity_general}. For an exact block-encoding of $e^{-\beta\hat H}$ with subnormalization $\alpha(\beta)$, that equation gives $P_{\mathrm{succ}}(\beta)\,F_g(\beta) = \gamma_0\, e^{-2\beta E_0}/\alpha(\beta)^2$. In the commuting termwise-LCU construction, the product of the individual Pauli-term blocks is exact and has $\alpha_{\mathrm{LCU}}(\beta) = e^{\beta W}$. This gives the product form $\gamma_0\, e^{-2\beta(W+E_0)}$ in Eq.~\eqref{eq:PF_identity}. Proposition~\ref{thm:exact_identity} states the individual spectral formulas~\eqref{eq:P_LCU}--\eqref{eq:F_g} for $P_{\mathrm{LCU}}(\beta)$ and $F_g(\beta)$, obtained by direct eigenbasis expansion in Appendix~\ref{app:exact_identity}. These individual formulas support the state-dependent $\beta^\star$ rule in Corollary~\ref{cor:beta_star}; the role of pairwise commutativity in removing Trotter error is revisited in Remark~(iii) below.

\begin{proposition}[Exact identity]
\label{thm:exact_identity}
For any commuting Hamiltonian of the form above, any initial state $|\psi_0\rangle$, and any $\beta \ge 0$ (with $\beta = 0$ the identity block),
\begin{align}
\label{eq:P_LCU}
P_{\mathrm{LCU}}(\beta) \;&=\; e^{-2\beta W} \sum_{k} |c_k|^2\, e^{-2\beta E_k}, \\
\label{eq:F_g}
F_g(\beta) \;&=\; \frac{\gamma_0\, e^{-2\beta E_0}}{\sum_{k} |c_k|^2\, e^{-2\beta E_k}}, \\
\label{eq:PF_identity}
P_{\mathrm{LCU}}(\beta)\, F_g(\beta) \;&=\; \gamma_0 \, e^{-2\beta(W + E_0)}.
\end{align}
\end{proposition}

Although elementary, this identity makes explicit the exact tradeoff between the LCU success probability and the ground-subspace fidelity.

Note that $W + E_0 \ge 0$ always holds: from $\|\hat H\| \le \sum_{\boldsymbol\mu} |x_{\boldsymbol\mu}| = W$ (triangle inequality with $\|\sigma_{\boldsymbol\mu}\|_{\mathrm{op}} = 1$) we obtain $E_0 \ge -W$, so the envelope $e^{-2\beta(W+E_0)} \in (0,1]$ for all $\beta \ge 0$ and the product $P_{\mathrm{LCU}}(\beta) F_g(\beta) \le \gamma_0 \le 1$ is automatic.

In Eq.~\eqref{eq:PF_identity}, the product of the LCU success probability and the ground-subspace fidelity decays purely exponentially in $\beta$, with rate $2(W + E_0)$ set by the Hamiltonian decomposition and spectrum, and prefactor $\gamma_0$ equal to the initial-state overlap with the ground subspace. The individual factors $P_{\mathrm{LCU}}$ and $F_g$ trade off against each other: for $\gamma_0 > 0$, larger $\beta$ drives $F_g \to 1$, and $P_{\mathrm{LCU}}$ follows the ground-state envelope $\gamma_0\, e^{-2\beta(W+E_0)}$, vanishing when $W+E_0 > 0$ and remaining at $\gamma_0$ when $W+E_0 = 0$. Their product is fixed up to a single exponential envelope. This separation cleanly isolates the two sources of cost: the exponent $W + E_0$ and the initial-state overlap $\gamma_0$.

Three practical caveats are worth stating immediately, before the corollaries. First, $W = \sum_{\boldsymbol\mu \ne 0}|x_{\boldsymbol\mu}|$ is a property of the \emph{chosen} Pauli decomposition, not of the abstract operator $\hat H$. Merging duplicate Pauli strings and pulling identity shifts out of the LCU (absorbing them into a classical scalar $e^{-\beta c}$) does not increase $W$, and can strictly reduce it when opposite-sign contributions cancel or when nonzero identity components are present, without changing the fidelity. This is part of the algorithm design. Second, evaluating $\beta^\star$ below requires lower bounds on $\gamma_0$ and the spectral gap $\Delta$. Neither is available for free on a generic problem instance: classical preprocessing, warm-starting, or auxiliary quantum subroutines for overlap and gap estimation are needed to operationalize the rule. The identity does not remove this dependence; it only makes it explicit. Third, the exponent $W+E_0$ is specific to the termwise-LCU subnormalization $\alpha_{\mathrm{LCU}}(\beta) = e^{\beta W}$. For a generic block-encoding with subnormalization $\alpha(\beta)$, the same calculation gives $P_{\mathrm{succ}}(\beta)\,F_g(\beta) = \gamma_0\, e^{-2\beta E_0}/\alpha(\beta)^2$, so a smaller achievable $\alpha(\beta)$ would reduce the exponent. The normalization lower bound is $\alpha(\beta) \ge \|e^{-\beta\hat H}\|_{\mathrm{op}} = e^{-\beta E_0}$; attaining this bound would remove the $\beta$-dependent decay of $P\,F_g$, but would require ground-energy information and an efficient corresponding block-encoding. QSVT-based alternatives are noted in Sec.~\ref{DISC}.

A gap-dependent lower bound on $F_g$ follows directly from Eq.~\eqref{eq:F_g}.

\begin{corollary}[Gap bound on fidelity]
\label{cor:gap_bound}
Let $\Delta := \min_{k \notin \mathcal G}(E_k - E_0) > 0$ be the spectral gap above the ground subspace. Then
\begin{align}
\label{eq:gap_bound}
F_g(\beta) \;\ge\; \frac{\gamma_0}{\gamma_0 + (1-\gamma_0)\, e^{-2\beta \Delta}},
\end{align}
which for $\gamma_0 > 0$ equals $\bigl[\, 1 + \tfrac{1 - \gamma_0}{\gamma_0}\, e^{-2\beta \Delta} \bigr]^{-1}$ and for $\gamma_0 = 0$ reduces to the trivial bound $F_g(\beta) \ge 0$.
\end{corollary}

% The bound of Corollary~\ref{cor:gap_bound} follows from the eigenbasis form~\eqref{eq:F_g} alone and therefore applies to the idealized ITE state $e^{-\beta\hat H}|\psi_0\rangle/\|e^{-\beta\hat H}|\psi_0\rangle\|$ as well; related gap-based fidelity and evolution-time scaling relations appear in Yi \textit{et al.}~\cite{yi2025probabilistic}. The role of Proposition~\ref{thm:exact_identity} is to relate it to the LCU success probability $P_{\mathrm{LCU}}(\beta^\star)$ via Eq.~\eqref{eq:PF_identity}.
The bound of Corollary~\ref{cor:gap_bound} follows from the eigenbasis form~\eqref{eq:F_g} alone and therefore applies to the idealized ITE state $e^{-\beta\hat H}|\psi_0\rangle/\|e^{-\beta\hat H}|\psi_0\rangle\|$ as well; related gap-based fidelity and evolution-time scaling relations appear in Yi \textit{et al.}~\cite{yi2025probabilistic}. In contrast to such one-sided bounds on $F_g$, Proposition~\ref{thm:exact_identity} provides an exact identity for the joint quantity $P_{\mathrm{LCU}}(\beta)\,F_g(\beta)$, directly coupling the fidelity to the post-selection success probability that determines amplification cost via Eq.~\eqref{eq:PF_identity}.

Inverting Eq.~\eqref{eq:gap_bound} gives a sufficient choice of $\beta$ for any target fidelity $\bar F \in (0, 1)$; the nontrivial regime is $\bar F \in (\gamma_0, 1)$, while $\bar F \le \gamma_0$ is already satisfied at $\beta = 0$.

\begin{corollary}[Gap-based sufficient $\beta$ for target fidelity]
\label{cor:beta_star}
Assume $0 < \gamma_0 < 1$, $\Delta > 0$, and $\gamma_0 < \bar F < 1$. (For $\gamma_0 \ge \bar F$, $\beta = 0$ already suffices.) To guarantee $F_g(\beta) \ge \bar F$, it suffices to take
\begin{align}
\label{eq:beta_star}
\beta \;\ge\; \beta^\star(\bar F) \; := \; \max\!\left\{ 0, \;\; \frac{1}{2\Delta}\, \log\!\left( \frac{\bar F\,(1 - \gamma_0)}{\gamma_0\,(1 - \bar F)} \right) \right\}.
\end{align}
\end{corollary}

The logarithm is positive precisely in the nontrivial regime $\gamma_0 < \bar F < 1$. For $\bar F \le \gamma_0$ the initial state already satisfies the target and $\beta = 0$ suffices. For $\bar F = 1$ the threshold is unattainable at any finite $\beta$ whenever $\gamma_0 < 1$. The degenerate case $\gamma_0 = 0$ is excluded from the formula. If $\gamma_0 = 0$ and $\bar F > 0$, no finite $\beta$ achieves $F_g(\beta) \ge \bar F$ (equivalently, $\beta^\star(\bar F) = +\infty$). The identity $P_{\mathrm{LCU}} F_g = \gamma_0 e^{-2\beta(W+E_0)} = 0$ then reflects the information-theoretic obstruction of an initial state with no ground-subspace overlap.

Combining Eq.~\eqref{eq:beta_star} with Eq.~\eqref{eq:PF_identity} and using $F_g(\beta^\star) \le 1$ yields a lower bound on the LCU success probability at the fidelity threshold:
\begin{align}
\label{eq:P_at_beta_star}
P_{\mathrm{LCU}}(\beta^\star) \;\ge\; \gamma_0\, e^{-2\beta^\star (W + E_0)},
\end{align}
A matching upper bound $P_{\mathrm{LCU}}(\beta^\star) \le (\gamma_0/\bar F)\, e^{-2\beta^\star (W+E_0)}$ follows from $F_g(\beta^\star) \ge \bar F$. The two bounds match in exponential dependence and differ only by the factor $1/\bar F$. Setting $\lambda_\star = \gamma_0\, e^{-2\beta^\star(W+E_0)} \le P_{\mathrm{LCU}}(\beta^\star)$ and substituting into the FPAA query bound $L = \mathcal O(\log(2/\delta)/\sqrt{\lambda_\star})$~\cite{yoder2014fixed} yields the FPAA query complexity, up to the additional cost of reflections, oracles, and initial-state preparation. Writing this explicitly, at the fidelity threshold the FPAA query complexity scales as
\begin{align}
\label{eq:end_to_end_cost}
L \;=\; \mathcal O\!\left(\frac{\log(2/\delta)\, e^{\beta^\star (W+E_0)}}{\sqrt{\gamma_0}}\right).
\end{align}
The exponent $W+E_0$, the threshold $\beta^\star$ of Corollary~\ref{cor:beta_star}, and the initial-state overlap $\gamma_0$ jointly determine this query complexity. Appendix~\ref{app:cnot_complexity} combines this with the reflection and LCU gate counts of each FPAA iteration to obtain the total CNOT count. The spectral expression~\eqref{eq:P_LCU} is a state-dependent input that is at least as tight as the state-independent termwise LCU bound~\cite{childs2012Hamiltonian}, and generically strictly tighter.

\paragraph*{Regimes of $\gamma_0$.}
The initial overlap $\gamma_0$ controls the required $\beta$. (i) In the \emph{warm-start} regime $\gamma_0 \gtrsim 1/2$ with target fidelity bounded away from $1$, the logarithm in Eq.~\eqref{eq:beta_star} is $\mathcal O(1)$; when $\Delta$ is constant, a small $\beta$ already gives high fidelity. (ii) For the diagonal QUBO/HUBO case treated in this paper with uniform-superposition initial state $|\psi_0\rangle = |+\rangle^{\otimes n}$, one has $\gamma_0 = |\mathcal G|/2^n$. If $\gamma_0 \ge 1/\mathrm{poly}(n)$ and $\Delta = \Omega(1/\mathrm{poly}(n))$, then for any fixed target fidelity $\bar F<1$, $\beta^\star = \mathcal O(\Delta^{-1}\log n)$. (iii) In the \emph{adversarial} regime $\gamma_0 \to 0$, $\beta^\star$ diverges logarithmically. The identity~\eqref{eq:PF_identity} then shows that no choice of $\beta$ can simultaneously keep both $P_{\mathrm{LCU}}$ and $F_g$ large. This is an information-theoretic barrier shared by any post-selection-based ITE, not a deficiency unique to the present method.

\paragraph*{Relation to the loose product bound.}
The state-independent termwise LCU bound~\cite{childs2012Hamiltonian} (see Appendix~\ref{app:LCU}) takes the form
\begin{equation}
P_{\mathrm{LCU}} \ge \prod_{\boldsymbol\mu} \left[1 - \frac{4\coth(\beta|x_{\boldsymbol\mu}|)}{(\coth(\beta|x_{\boldsymbol\mu}|) + 1)^2}\right],
\end{equation}
which simplifies, via the identity $1 - 4\coth(y)/(\coth(y)+1)^2 = e^{-4y}$, to $P_{\mathrm{LCU}} \ge e^{-4\beta W}$. The same worst-case bound follows directly from Eq.~\eqref{eq:P_LCU}: since $E_k \le \|\hat H\|_{\mathrm{op}} \le W$ and $\sum_k |c_k|^2 = 1$, one has $\sum_k |c_k|^2 e^{-2\beta E_k} \ge e^{-2\beta W}$, hence $P_{\mathrm{LCU}}(\beta) \ge e^{-4\beta W}$. The two expressions coincide in the worst case. The spectral form~\eqref{eq:P_LCU} is strictly tighter whenever the initial state concentrates on eigenstates with $E_k < W$, because it retains the actual energy distribution $\{|c_k|^2\}$ rather than a term-by-term worst-case product. We use Eq.~\eqref{eq:P_LCU} as the operating expression in the remainder of this paper.

\paragraph*{Applicability.}
The closed-form reciprocal $F_g(\beta) \ge [1 + ((1-\gamma_0)/\gamma_0)\,e^{-2\beta\Delta}]^{-1}$ of Corollary~\ref{cor:gap_bound} requires $\gamma_0 > 0$; the $\beta^\star$ rule of Corollary~\ref{cor:beta_star} additionally requires $\Delta > 0$. The bound Eq.~\eqref{eq:gap_bound} itself remains well defined at $\gamma_0 = 0$ and reduces there to the trivial statement $F_g(\beta) \ge 0$. For each fixed finite-dimensional instance with at least one excited level, $\Delta > 0$. The \emph{gapless} regime $\Delta \to 0$ refers either to an asymptotic family with $\Delta_n \to 0$ or, degenerately, to a Hamiltonian proportional to the identity. In either case $\beta^\star$ is not well defined, and a spectral gap assumption must be made at the problem-instance level. The endpoint cases are trivial: $\gamma_0 = 0$ is the adversarial regime discussed above, while $\gamma_0 = 1$ gives $F_g(\beta) = 1$ for all $\beta$, so one may set $\beta^\star = 0$. The $\beta^\star$ value of Corollary~\ref{cor:beta_star} places all excited weight at the gap energy $E_0 + \Delta$; for generic spectra it is therefore a conservative upper bound on the sufficient imaginary time, and any spectrum with first excited gap $\ge \Delta$ attains the stated target fidelity at $\beta \ge \beta^\star$.

\paragraph*{Remarks.}
Three features of the above identity deserve emphasis.

\emph{(i) Decomposition dependence of $W$.} The constant $W = \sum_{\boldsymbol\mu \ne 0} |x_{\boldsymbol\mu}|$ is not an invariant of the Hamiltonian $\hat H$. It is the $\ell_1$-norm of the particular Pauli decomposition fed to the LCU. Equivalent Pauli strings should be merged before forming the circuit, and any identity component $c\,I$ is best removed from the LCU and applied as a classical scalar $e^{-\beta c}$; including it inflates $W$ without changing $F_g$. This choice is part of the algorithm design, not of the problem instance.

\emph{(ii) Ground-space, not ground-state, preparation.} For a degenerate ground subspace $\mathcal G$ with $|\mathcal G| > 1$ and $\gamma_0 > 0$, the normalized post-selected state converges to $\Pi_{\mathcal G} |\psi_0\rangle / \sqrt{\gamma_0}$ as $\beta \to \infty$. This is the projection of the initial state onto $\mathcal G$, not any particular basis element of $\mathcal G$. The fidelity $F_g$ defined here measures total ground-space weight and is the appropriate figure of merit in this setting.

\emph{(iii) Commutativity and exactness.} All three identities are \emph{exact} under the standing assumption that the Pauli strings $\{\sigma_{\boldsymbol\mu}\}$ pairwise commute and that the LCU is implemented as an ideal block-encoding. No Trotter error enters because $e^{-\beta \hat H} = \prod_{\boldsymbol\mu} e^{-\beta x_{\boldsymbol\mu} \sigma_{\boldsymbol\mu}}$ is itself exact. For non-commuting Hamiltonians, an additional product-formula error would enter separately from the post-selection analysis given here.

\section{Numerical Demonstration}\label{POPND}

%-----------------------------------------------------------------------------------------------------------------
We validate the FinITE analysis of Sec.~\ref{FinITE} on two representative PUBO instances: a quadratic MAX-CUT QUBO instance and a cubic weighted HUBO instance. Proposition~\ref{thm:exact_identity} applies to any pairwise-commuting PUBO Hamiltonian; these examples test both the quadratic and higher-order settings discussed in Sec.~\ref{sec:background}. Sections~\ref{subsec:setup}--\ref{subsec:validation} apply the full LCU and FPAA procedure to a five-vertex MAX-CUT (QUBO) instance. We use Proposition~\ref{thm:exact_identity} and Corollary~\ref{cor:beta_star} on the graph to derive numerical predictions. These are compared against Qiskit \texttt{qasm\_simulator}~\cite{qiskit2024} results for two observables---the LCU success probability and the conditional ground-subspace fidelity---in two regimes: without FPAA, and under FPAA at varying query complexity $L$. Sec.~\ref{subsec:hubo_validation} extends the validation to a native HUBO instance combining cubic and quadratic Pauli terms. Using Qiskit's statevector simulator, we perform an exact identity check across three initial-state regimes spanning nearly an order of magnitude in $\gamma_0$. The HUBO subsection is restricted to verifying the identity of Proposition~\ref{thm:exact_identity}. Once a lower bound $\lambda \le P_{\mathrm{LCU}}(\beta)$ is supplied, the FPAA query formula $L = \mathcal O(\log(2/\delta)/\sqrt{\lambda})$ and amplification schedule do not depend on the Hamiltonian. However, the LCU success curve and LCU gate complexity remain Hamiltonian-dependent.

\subsection{Test graph and identity predictions}
\label{subsec:setup}

\begin{figure}[!htbp]
	\centering
	\includegraphics[width=7cm]{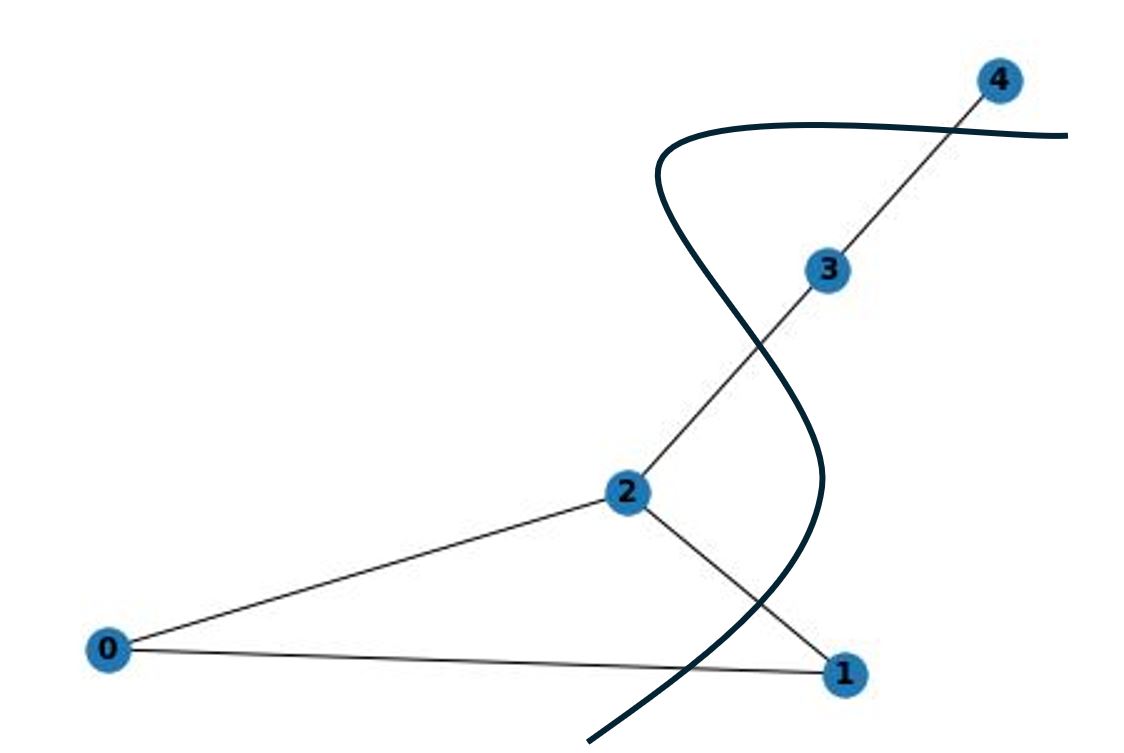}
	\caption{Five-vertex MAX-CUT test graph with $|E|=5$ edges. The curve indicates one of the three optimal cuts (six ground-state labelings counting partition complements); the graph is non-bipartite, so no cut separates every edge. The five edges $\{(0,1),(0,2),(1,2),(2,3),(3,4)\}$ combine a triangle on $\{0,1,2\}$ with a path $2$--$3$--$4$.}
	\label{target_graph}
\end{figure}

MAX-CUT is the problem of finding a partition of the vertices of an undirected graph $(V, E)$ into two sets that maximizes the number of cut edges. Encoding each vertex's side by a $\sigma_z$ eigenvalue $\pm 1$, the problem maps to finding the ground states of the Ising Hamiltonian $\hat H = \sum_{(i,j)\in E} \sigma_z^{(i)}\sigma_z^{(j)}$~\cite{hadfield2021on,farhi2014quantum,wecker2016training}. Each edge contributes $-1$ to $\hat H$ when cut and $+1$ when uncut. We use the five-vertex graph of Fig.~\ref{target_graph} as a small but non-bipartite instance for which Proposition~\ref{thm:exact_identity} gives closed-form predictions that can be compared against simulation at every $\beta$.

For this graph the Pauli decomposition has
\begin{equation}
W \;=\; \sum_{(i,j) \in E} |x_{ij}| \;=\; |E| \;=\; 5,
\end{equation}
since each edge contributes a unit-coefficient $\sigma_z^{(i)}\sigma_z^{(j)}$ term. Direct enumeration of the $2^5 = 32$ computational-basis states gives ground-state energy $E_0 = -3$ (maximum cut of $4$ out of $5$ edges) and ground-subspace degeneracy $|\mathcal G| = 6$ (three cuts of size four together with their $\mathbb Z_2$-complements). The lowest strictly excited energy is $-1$, so the spectral gap is $\Delta = \min_{k \notin \mathcal G}(E_k - E_0) = 2$. For the uniform initial state $|\psi_0\rangle = |+\rangle^{\otimes 5}$, the ground-space overlap is
\begin{equation}
\gamma_0 \;=\; \frac{|\mathcal G|}{2^5} \;=\; \frac{6}{32} \;=\; \frac{3}{16}.
\end{equation}

Substituting into the identity Eq.~\eqref{eq:PF_identity} gives the closed-form prediction
\begin{equation}
\label{eq:graph_prediction}
P_{\mathrm{LCU}}(\beta) \cdot F_g(\beta) \;=\; \frac{3}{16}\, e^{-4\beta}
\qquad \text{for every } \beta \ge 0.
\end{equation}
At $\beta = 0$ this recovers $1 \cdot (3/16) = 3/16$, matching the initial overlap $\gamma_0$. The sufficient-$\beta$ rule of Corollary~\ref{cor:beta_star} evaluates to
\begin{equation}
\label{eq:graph_beta_star}
\beta^\star(\bar F) \;=\; \frac{1}{4}\, \log\!\left( \frac{13\, \bar F}{3\,(1 - \bar F)} \right),
\end{equation}
giving $\beta^\star(0.5) \approx 0.37$, $\beta^\star(0.9) \approx 0.92$, and $\beta^\star(0.98) \approx 1.34$. All of these fall within the simulated range $\beta \in [0, 2]$, and we check them against simulation next.

\subsection{MAX-CUT: shot-based validation}
\label{subsec:validation}

We now test the 5-vertex MAX-CUT instance of Sec.~\ref{subsec:setup} with shot-based simulations, comparing against its closed-form predictions. For MAX-CUT, the termwise-LCU operator factorizes as $\prod_{(i,j)\in E}[\cosh(\beta)I - \sinh(\beta)\sigma_z^{(i)}\sigma_z^{(j)}]/e^\beta$ (for $\beta>0$ each edge contributes $\kappa = \coth(\beta)$ in the notation of Sec.~\ref{imp_qc}, and the denominator $e^\beta = \cosh(\beta) + \sinh(\beta)$ normalizes each block to operator norm $1$; at $\beta = 0$ the block reduces to the identity). Simulations use the initial state $|\psi_0\rangle = |+\rangle^{\otimes 5}$ and $\beta \in [0, 2]$ with step $0.001$.

\begin{figure}[!htbp]
\centering
\includegraphics[width=10cm]{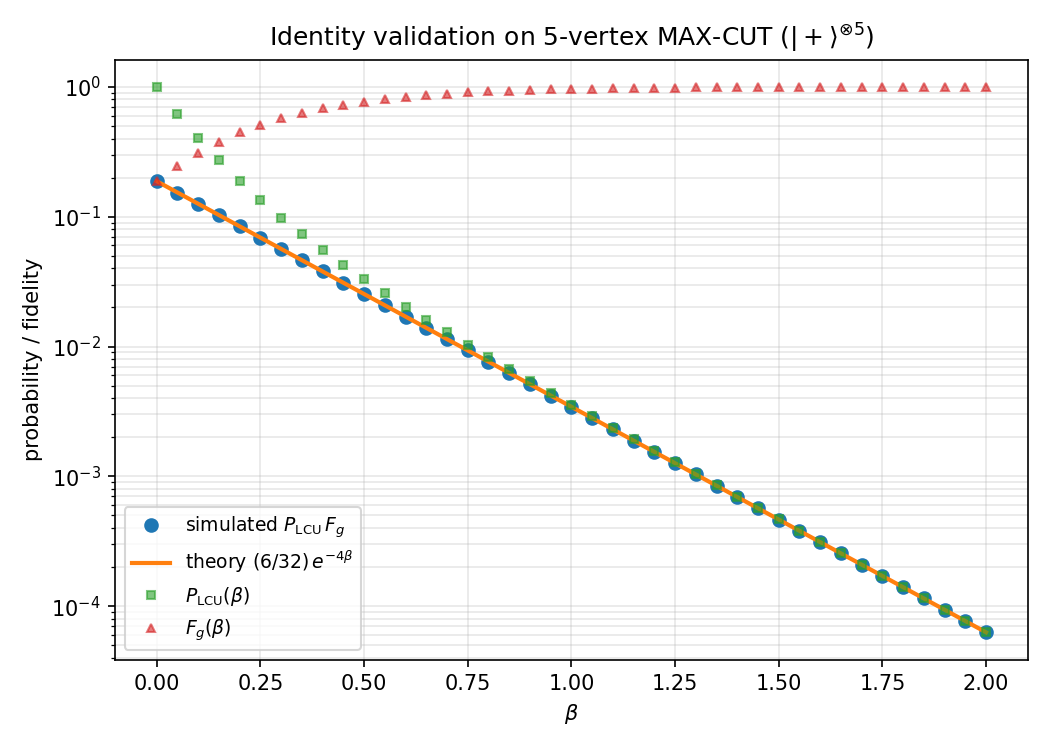}
\caption{Identity validation on the 5-vertex MAX-CUT instance with $|\psi_0\rangle = |+\rangle^{\otimes 5}$. Green circles: simulated $P_{\mathrm{LCU}}(\beta)\,F_g(\beta)$; solid line: closed-form prediction $(3/16)\,e^{-4\beta}$. Blue squares: $P_{\mathrm{LCU}}(\beta)$; red triangles: $F_g(\beta)$. Maximum relative error is $1.3 \times 10^{-14}$, near the double-precision roundoff level.}
\label{fig:maxcut_identity}
\end{figure}

We first verify the identity of Proposition~\ref{thm:exact_identity} at statevector precision: Fig.~\ref{fig:maxcut_identity} shows that the simulated product $P_{\mathrm{LCU}}(\beta)\,F_g(\beta)$ matches the closed-form prediction $(3/16)\,e^{-4\beta}$ with maximum relative error $1.3\times 10^{-14}$, near the double-precision roundoff level.

\begin{figure}[!htbp]
\centering
\includegraphics[width=\textwidth]{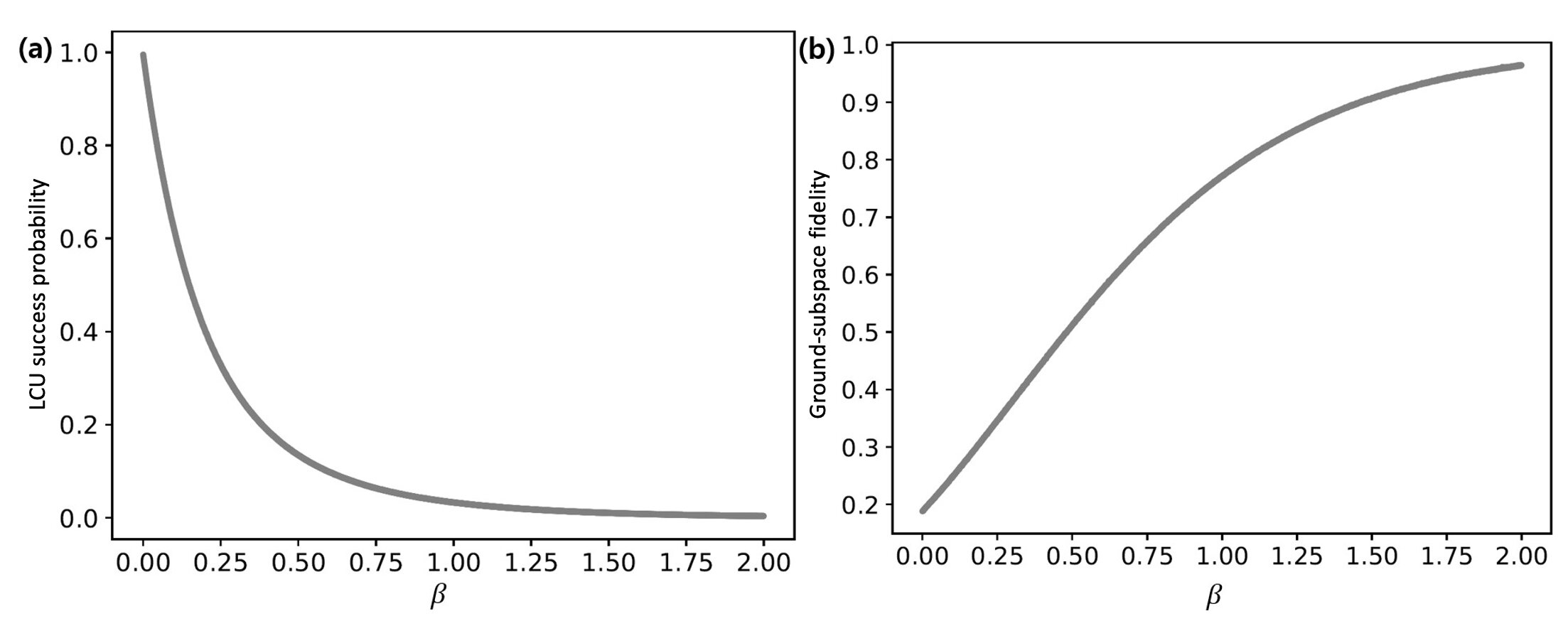}
\caption{Simulation results of MAX-CUT without amplitude amplification. (a) LCU success probability $P_{\mathrm{LCU}}(\beta)$. (b) Conditional ground-subspace fidelity $F_g(\beta)$, the probability of finding the system in the ground subspace given that the LCU succeeded.}
\label{without_am}
\end{figure}

We then turn to Qiskit \texttt{qasm\_simulator} shot-based measurements. Fig.~\ref{without_am} tests the two factors of the prediction Eq.~\eqref{eq:graph_prediction} separately, measuring (a) $P_{\mathrm{LCU}}(\beta)$ and (b) $F_g(\beta)$ (the conditional ground-subspace fidelity given LCU success) as $\beta$ is swept from $0$ to $2$. Shot counts range from $10^3$ to $10^5$ per $\beta$ value; at this 5-qubit system size, the reported curves are consistent across this range.

Panel (a) shows $P_{\mathrm{LCU}}(\beta)$ falling rapidly from $1$ at $\beta = 0$ and decaying toward zero as $\beta$ grows, consistent with the $e^{-2(W+E_0)\beta} = e^{-4\beta}$ envelope on the product $P_{\mathrm{LCU}}\,F_g$ predicted by Eq.~\eqref{eq:graph_prediction} for this non-bipartite instance with $W+E_0 = 2 > 0$; the individual factor $P_{\mathrm{LCU}}$ is further modulated by the state-dependent denominator of Eq.~\eqref{eq:P_LCU}, and stays strictly positive at every finite $\beta$. Panel (b) shows $F_g(\beta)$ rising monotonically as $\beta$ grows. Combining the two panels, the product $P_{\mathrm{LCU}}(\beta)\,F_g(\beta)$ follows the single exponential $(3/16)\,e^{-4\beta}$ of Eq.~\eqref{eq:graph_prediction}: the rising $F_g$ and falling $P_{\mathrm{LCU}}$ are the two sides of this single envelope, confirming that ITE is operating correctly and exposing the trade-off that motivates FPAA.

\begin{figure}[!htbp]
\centering
\includegraphics[width=\textwidth]{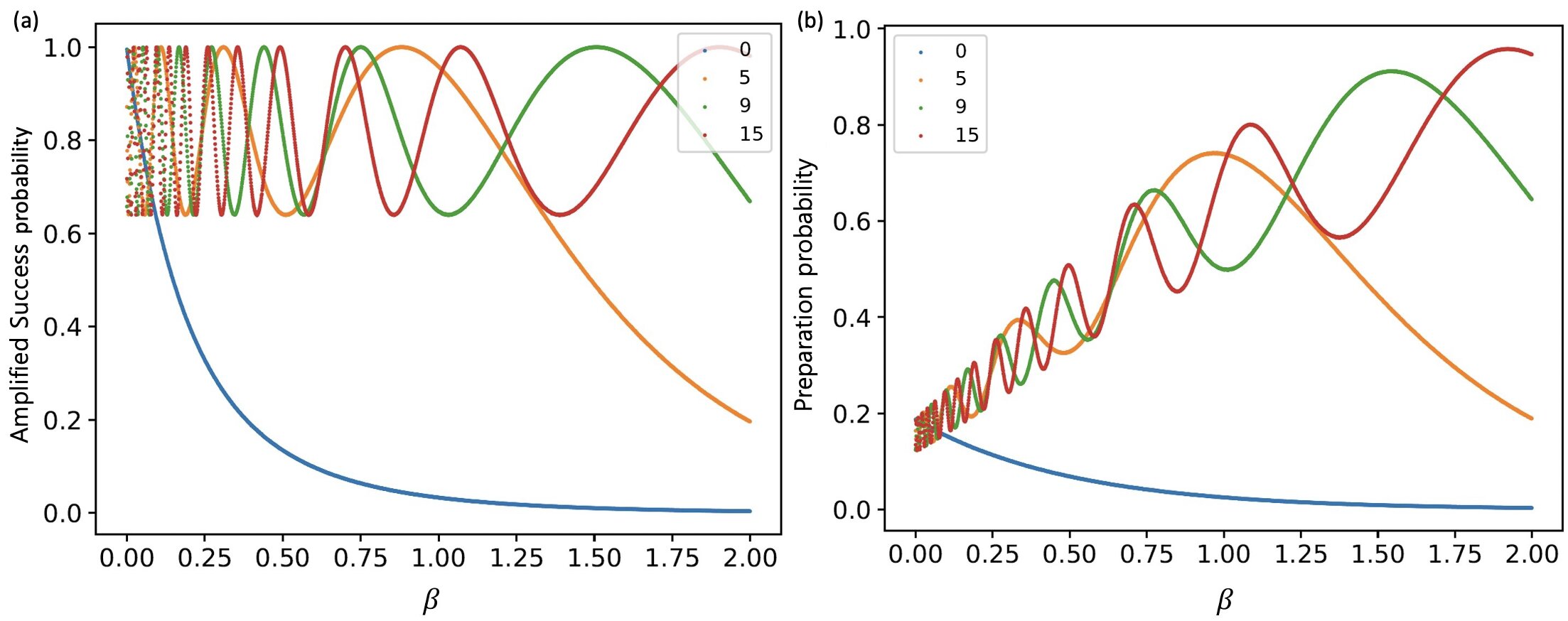}
\caption{Simulation results with varying query complexities $L$. (a) FPAA-amplified LCU success probability $P_{\mathrm{amp}}^{(L)}(\beta)$. (b) Ground-subspace preparation probability $P_g^{(L)}(\beta) = P_{\mathrm{amp}}^{(L)}(\beta)\,F_g(\beta)$ for different $L$, i.e., the joint probability of amplified LCU success and being in the ground subspace.}
\label{with_am_various_L}
\end{figure}

Fig.~\ref{with_am_various_L} turns on FPAA at varying $\beta$ across multiple query complexities $L$ and tracks, as $\beta$ is swept from $0$ to $2$, both (a) the amplified LCU success probability $P_{\mathrm{amp}}^{(L)}(\beta)$ and (b) the ground-subspace preparation probability $P_g^{(L)}(\beta) = P_{\mathrm{amp}}^{(L)}(\beta)\, F_g(\beta)$, defined as the joint probability of amplified LCU success and being in the ground subspace, where $F_g(\beta)$ is the conditional ground-subspace fidelity of the normalized post-selected state. FPAA amplifies $P_{\mathrm{LCU}}$ but leaves this conditional fidelity unchanged. The query complexity $L$ is a tunable parameter: larger $L$ widens the $\beta$-range over which FPAA can keep the amplified success probability near unity despite the exponential decay of the unamplified $P_{\mathrm{LCU}}(\beta)$.

Fig.~\ref{with_am_various_L}(a) shows the amplified success probability for $L = 0, 5, 9, 15$. $L = 0$ reproduces Fig.~\ref{without_am}(a). For $L > 0$, the curve oscillates with $\beta$. These oscillations are the Chebyshev-like response of the FPAA sequence~\cite{yoder2014fixed}: at fixed $L$, the FPAA polynomial yields an output close to unity when the unamplified success probability $P_{\mathrm{LCU}}(\beta)$ exceeds an $L$-dependent threshold and decays below it. As $\beta$ increases, $P_{\mathrm{LCU}}(\beta)$ crosses this threshold, producing the observed oscillations. For $L = 5$ the amplification is already insufficient once $\beta$ passes $\beta^\star(0.9) \approx 0.92$, and the curve decays toward zero for $\beta \gtrsim 1$.

Fig.~\ref{with_am_various_L}(b) shows the corresponding ground-subspace preparation probability $P_g^{(L)}(\beta)$. For $L = 0$, $P_g^{(0)}(\beta) = P_{\mathrm{LCU}}(\beta)\,F_g(\beta)$, the fixed envelope of Eq.~\eqref{eq:graph_prediction}, starting at $\gamma_0 = 3/16$ and decaying monotonically. For $L > 0$, the oscillating amplified success probability in panel (a) combines with the monotone growth of $F_g(\beta)$ in Fig.~\ref{without_am}(b), producing a $P_g^{(L)}(\beta)$ that oscillates between low values and values approaching $F_g(\beta)$. Since $P_g^{(L)}(\beta) \le F_g(\beta)$ at any $L$, a near-unity $P_g^{(L)}$ requires both successful amplification and $\beta$ large enough that $F_g(\beta)$ itself is close to one. $L = 15$ supplies enough amplification for $P_g^{(L)}$ to approach unity once $\beta$ exceeds $\beta^\star(0.98) \approx 1.34$ within the simulated range.

FPAA cannot push $P_g^{(L)}(\beta)$ past a target threshold while $F_g(\beta)$ itself remains below that threshold, since FPAA boosts $P_{\mathrm{LCU}}$ but not $F_g$. Above the corresponding $F_g$-crossing, the required query complexity scales as $L \sim 1/\sqrt{P_{\mathrm{LCU}}(\beta)} = e^{2\beta}\sqrt{F_g(\beta)/\gamma_0}$ using Eq.~\eqref{eq:graph_prediction}, so once $F_g(\beta)$ is close to one the dominant scaling is exponential in $\beta$ with rate $2$ on this instance.

Taken together, the statevector-level identity check of Fig.~\ref{fig:maxcut_identity} and the shot-based analysis of Figs.~\ref{without_am} and~\ref{with_am_various_L} validate the identity of Sec.~\ref{FinITE} on this instance. In Fig.~\ref{with_am_various_L}(b) at $L = 15$, $P_g^{(L)}$ approaches unity before $\beta^\star(0.98) \approx 1.34$, and the identity product $P_{\mathrm{LCU}}(\beta) F_g(\beta) = (3/16)\,e^{-4\beta}$ organizes the simultaneously decreasing $P_{\mathrm{LCU}}$ and increasing $F_g$ in Fig.~\ref{without_am} into a single fixed exponential curve.

\subsection{Native HUBO: weighted hypergraph instance}
\label{subsec:hubo_validation}

We extend the validation beyond QUBO to a native HUBO Hamiltonian. The identity envelope $\gamma_0\,e^{-2\beta(W+E_0)}$ is linear in $\gamma_0$, so raising $\gamma_0$ through a warm start directly shifts the envelope upward; we therefore sweep three initial-state regimes to probe both this prefactor structure and the warm-start effect. The Hamiltonian contains cubic and quadratic Pauli-$Z$ terms with non-uniform coefficients:
\begin{align}
\hat H_{\mathrm{HUBO}} \;=\;&
 2.0\,\sigma_z^{(0)}\sigma_z^{(1)}\sigma_z^{(2)}
 + 1.5\,\sigma_z^{(1)}\sigma_z^{(2)}\sigma_z^{(3)} \nonumber \\
&+ 2.0\,\sigma_z^{(4)}\sigma_z^{(5)}\sigma_z^{(6)}
  + 1.5\,\sigma_z^{(5)}\sigma_z^{(6)}\sigma_z^{(7)} \nonumber \\
&+ 1.0\,\sigma_z^{(0)}\sigma_z^{(4)}
  + 1.0\,\sigma_z^{(3)}\sigma_z^{(7)} \nonumber \\
&+ 1.2\,\sigma_z^{(0)}\sigma_z^{(3)}
  + 1.2\,\sigma_z^{(4)}\sigma_z^{(7)}.
\label{eq:hubo_instance}
\end{align}
This 8-qubit, $M = 8$-term instance is a native weighted HUBO Hamiltonian with cubic Pauli-$Z$ parity terms and frustrated quadratic bridges. Structurally, this is a parity-HUBO instance (3-local $\sigma_z$ parity terms coupled by 2-local bridges), analogous to small MAX-3-XORSAT / MAX-3-SAT instances in the combinatorial optimization literature. Direct enumeration of the $2^8 = 256$ computational-basis states gives $W = 11.4$, $E_0 = -6.6$, and $|\mathcal G| = 8$. The spectral gap is $\Delta = \min_{k \notin \mathcal G}(E_k - E_0) = 1.2$, realized by the coordinated bit-flip pair $(b_3, b_7)$ that satisfies both frustrated quadratic bridges at the cost of violating two cubic terms. For the uniform state $|\psi_0\rangle = |+\rangle^{\otimes 8}$, $\gamma_0 = |\mathcal G|/2^8 = 1/32$, and the identity Eq.~\eqref{eq:PF_identity} predicts
\begin{equation}
P_{\mathrm{LCU}}(\beta)\, F_g(\beta) \;=\; \frac{1}{32}\, e^{-9.6\,\beta}
\qquad \text{for every } \beta \ge 0.
\label{eq:hubo_prediction}
\end{equation}

We implement the termwise-LCU circuit of Sec.~\ref{imp_qc} with $M = 8$ ancillae, one for each Pauli term. We run Qiskit's statevector simulator~\cite{qiskit2024} (exact) over $\beta \in [0, 3]$ for three initial-state regimes spanning nearly an order of magnitude in $\gamma_0$. The first is the uniform state $|+\rangle^{\otimes 8}$ ($\gamma_0 \approx 0.031$). The other two are product warm starts biased toward a chosen ground bitstring $g_\star$, in which each qubit $q$ is prepared as $\sqrt{p}\,|g_\star[q]\rangle + \sqrt{1-p}\,|1 \oplus g_\star[q]\rangle$ via a single $R_y$ rotation; the resulting overlap is $\gamma_0(p) = \sum_{g \in \mathcal G} p^{n-d(g,g_\star)}(1-p)^{d(g,g_\star)}$ with Hamming distance $d$ to $g_\star$. Here $g_\star$ is obtained by direct enumeration of the $2^8$ basis states to construct a controlled warm-start demonstrator; in larger instances, heuristic candidate bitstrings would be used instead (Sec.~\ref{DISC}). We use $p = 0.60$ ($\gamma_0 \approx 0.041$) and $p = 0.85$ ($\gamma_0 \approx 0.29$).

\begin{figure}[!htbp]
\centering
\includegraphics[width=10cm]{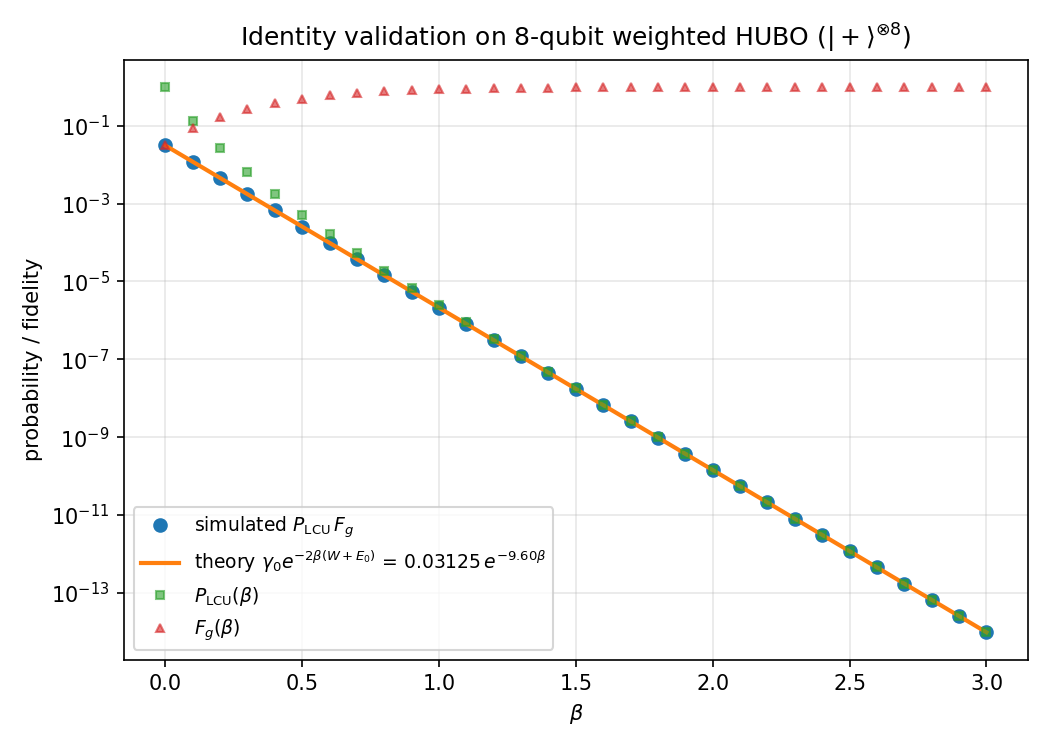}
\caption{Identity check on the weighted HUBO instance Eq.~\eqref{eq:hubo_instance} with uniform initial state $|+\rangle^{\otimes 8}$. Green circles: simulated $P_{\mathrm{LCU}}(\beta) F_g(\beta)$; solid line: closed-form prediction $(1/32)\,e^{-9.6\beta}$. Blue squares: $P_{\mathrm{LCU}}(\beta)$; red triangles: $F_g(\beta)$. Maximum relative error across the swept $\beta$ is $1.2 \times 10^{-12}$, at the double-precision floating-point limit.}
\label{fig:hubo_identity}
\end{figure}

Fig.~\ref{fig:hubo_identity} displays the identity check for the uniform regime: the simulated product matches the closed-form prediction across $\beta \in [0, 3]$ with maximum relative error $1.2 \times 10^{-12}$. Because the statevector simulator uses exact expectation values rather than finite-shot sampling, no shot noise enters; the agreement validates Eq.~\eqref{eq:PF_identity} as an equality rather than an approximation.

\begin{figure}[!htbp]
\centering
\includegraphics[width=10cm]{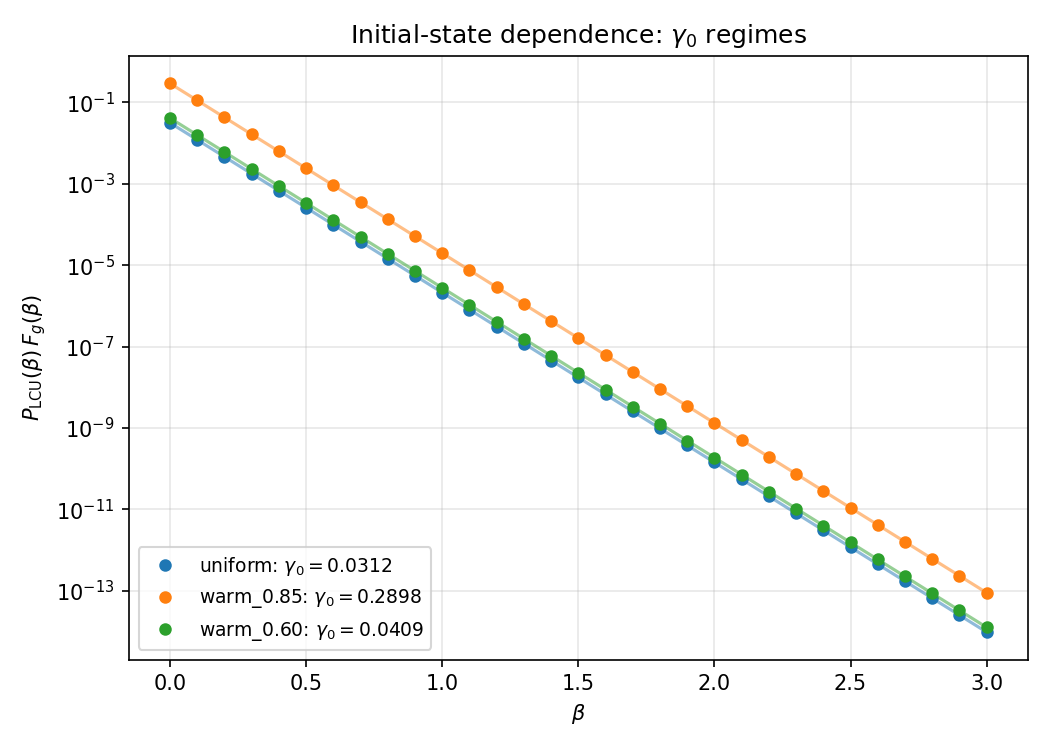}
\caption{Initial-state dependence on the HUBO Hamiltonian Eq.~\eqref{eq:hubo_instance}. The three regimes---uniform ($\gamma_0 \approx 0.031$), warm start at $p = 0.60$ ($\gamma_0 \approx 0.041$), and warm start at $p = 0.85$ ($\gamma_0 \approx 0.29$)---produce parallel envelopes $\gamma_0\,e^{-9.6\beta}$ with common slope $-2(W+E_0) = -9.6$; the prefactor is the initial overlap $\gamma_0$.}
\label{fig:hubo_gamma_regimes}
\end{figure}

Fig.~\ref{fig:hubo_gamma_regimes} sweeps the initial state. All three regimes follow the theory envelope $\gamma_0\,e^{-9.6\beta}$: the exponent $W + E_0 = 4.8$ is common across regimes, while $\gamma_0$ scales the prefactor by roughly an order of magnitude between uniform and strong warm-start. This separation is the one implied by Proposition~\ref{thm:exact_identity} and illustrates the role of $\gamma_0$ as the initial-state bottleneck in Corollary~\ref{cor:beta_star}. Equivalently, a warm start raises the envelope ceiling by the $\gamma_0$ ratio, shrinking the $\beta$ required to reach any given target fidelity.

This HUBO validation confirms Proposition~\ref{thm:exact_identity} at statevector precision for commuting Hamiltonians with mixed cubic and quadratic Pauli terms, without quadratization. The FPAA query formula $L = \mathcal O(\log(2/\delta)/\sqrt{\lambda})$ and amplification schedule from the MAX-CUT demonstration of Sec.~\ref{subsec:validation} carry over unchanged, since FPAA (Sec.~\ref{FinITE}, Stage 2) acts on the LCU-success subspace. Only the success curve $P_{\mathrm{LCU}}(\beta)$ and the LCU gate complexity are Hamiltonian-dependent; these follow from the same analysis, so we do not repeat the full LCU+FPAA procedure here.

\section{Discussion and Conclusion}\label{DISC}

FinITE is an LCU-based realization of imaginary-time evolution for pairwise-commuting PUBO Hamiltonians. It avoids taking $\beta \to \infty$, where the termwise-LCU operator converges to the projector onto the states that are simultaneously ground states of every individual Pauli term in $\hat H$. For typical commuting PUBO Hamiltonians no such simultaneous ground state exists and the limiting operator is zero; when it does exist (as for bipartite MAX-CUT), it often spans only an exponentially small subspace, so the LCU success probability vanishes. Instead, FinITE uses a finite value of $\beta$ chosen analytically.

The reduced LCU success probability at finite $\beta$ is compensated by FPAA~\cite{yoder2014fixed}. Its main consequence is a finite-$\beta$ regime in which fidelity, LCU success probability, and amplification cost can be related analytically. The identity $P_{\mathrm{LCU}}(\beta) F_g(\beta) = \gamma_0 e^{-2\beta(W + E_0)}$ shows that the cost separates into an exponent $W + E_0$ and an initial-state overlap $\gamma_0$, replacing term-by-term product bounds with a state-dependent closed form. This identity, valid for any pairwise-commuting PUBO Hamiltonian and any initial state, enables a closed-form selection rule $\beta^\star(\bar F)$ for a prescribed target fidelity under the stated spectral information. The construction applies natively to cubic and higher-order Pauli interactions without quadratization, trading additional FPAA queries for LCU success and fidelity guarantees under the pairwise-commuting PUBO assumption.

Two simulations verify the identity on representative commuting PUBO instances. For a five-vertex MAX-CUT (QUBO) with $|\psi_0\rangle = |+\rangle^{\otimes 5}$, full LCU+FPAA simulation matches the closed-form prediction $(3/16)\,e^{-4\beta}$. In Fig.~\ref{with_am_various_L}(b) at $L = 15$, the ground-subspace preparation probability approaches unity once $\beta \gtrsim \beta^\star(0.98) \approx 1.34$, the sufficient $\beta$ from Corollary~\ref{cor:beta_star}. For an 8-qubit weighted HUBO instance with mixed cubic and quadratic Pauli terms, statevector simulation confirms the identity to $\sim 10^{-12}$ relative error across three initial-state regimes (Figs.~\ref{fig:hubo_identity}, \ref{fig:hubo_gamma_regimes}). The initial-state sweep illustrates the role of $\gamma_0$ as a controllable prefactor in the identity envelope: a product warm start raises $\gamma_0$ by nearly an order of magnitude over the uniform initial state, while the $W + E_0$ exponent remains fixed.

Two directions merit further investigation. First, proof-of-principle demonstrations on small problem instances can validate FinITE on current QPU platforms, providing empirical calibration of the success-probability and fidelity curves at a scale where the $M$-ancilla and FPAA overheads remain tractable. Second, the role of $\gamma_0$ as a controllable prefactor invites problem-specific warm-start strategies in which classical heuristics (e.g., greedy or local-search routines on the cost function) generate candidate ground bitstrings used to prepare biased initial states; this goes beyond the present demonstration, which uses warm starts biased toward a known ground bitstring to illustrate the $\gamma_0$ scaling.

Together, these results identify a tunable-$\beta$ ITE primitive for commuting PUBO Hamiltonians, with explicit links among fidelity, LCU success probability, and amplification depth.

%------------------------------------------------------------------------------------
\section*{Acknowledgment}
%------------------------------------------------------------------------------------
This work received partial support from the following sources:
(1) Basic Science Research Program through the National Research Foundation of Korea (NRF), funded by the Ministry of Science and ICT (RS-2023-NR068116, RS-2025-03532992).
(2) Institute for Information \& Communications Technology Promotion (IITP) grant funded by the Korea government (MSIP) (No.\ 2019-0-00003), which focuses on the research and development of core technologies for programming, running, implementing, and validating fault-tolerant quantum computing systems.
(3) Korean ARPA-H Project via the Korea Health Industry Development Institute (KHIDI); Ministry of Health and Welfare, Republic of Korea (RS-2025-25456722).
(4) Yonsei University Research Fund under project number 2025-22-0140.

\medskip
\noindent\textbf{Funding}\\
National Research Foundation of Korea, Institute for Information \& Communications Technology Promotion, Korea Health Industry Development Institute, and Yonsei University.

\medskip
\noindent\textbf{Ethics approval and consent to participate}\\
Not applicable.

\bibliographystyle{quantum}
\bibliography{imaginary}

%===============================================================
% APPENDICES
%===============================================================

\appendix

\section{Notation}
\label{app:notation}

Table~\ref{tab:notation} collects the symbols used throughout the paper, with the section or equation where each is first introduced.

\begin{table}[h!]
\caption{Notation.}
\label{tab:notation}
\centering
\small
\begin{tabular}{ll}
\hline
Symbol & Meaning \\ \hline
$M$ & number of Pauli terms in $\hat H = \sum_{\boldsymbol\mu} x_{\boldsymbol\mu} \sigma_{\boldsymbol\mu}$ \\
$\hat H$ & PUBO Hamiltonian, pairwise-commuting in this work \\
$x_{\boldsymbol\mu} \in \mathbb R$ & real Pauli-decomposition coefficient \\
$\sigma_{\boldsymbol\mu}$ & $n$-qubit Pauli-$Z$ string \\
$W$ & $\sum_{\boldsymbol\mu \ne 0} |x_{\boldsymbol\mu}|$, decomposition-dependent $\ell_1$-norm (non-identity terms) \\
$|\psi_0\rangle$ & normalized initial state, $\sum_k |c_k|^2 = 1$ \\
$\{|E_k\rangle\}$ & orthonormal common eigenbasis, $\hat H|E_k\rangle = E_k|E_k\rangle$ \\
$c_k$ & amplitude $\langle E_k|\psi_0\rangle$ \\
$E_0$ & ground-state energy \\
$\mathcal G$ & set of ground-state indices $\{k : E_k = E_0\}$ \\
$\Pi_{\mathcal G}$ & projector $\sum_{k \in \mathcal G}|E_k\rangle\langle E_k|$ onto $\mathcal G$ \\
$\gamma_0$ & $\langle\psi_0|\Pi_{\mathcal G}|\psi_0\rangle = \sum_{k \in \mathcal G}|c_k|^2$ \\
$\Delta$ & spectral gap $\min_{k \notin \mathcal G}(E_k - E_0)$ \\
$m(\beta)$ & LCU scaling factor $e^{-\beta W}$ \\
$\kappa_{\boldsymbol\mu}$ & LCU coefficient $\coth(|\beta x_{\boldsymbol\mu}|) \ge 0$ for term $\boldsymbol\mu$ \\
$P_{\mathrm{LCU}}(\beta)$ & LCU success probability (Prop.~\ref{thm:exact_identity}) \\
$F_g(\beta)$ & conditional ground-subspace fidelity given LCU success \\
$\bar F$ & target fidelity, $\bar F \in (0,1)$ \\
$\beta^\star(\bar F)$ & sufficient $\beta$ for $F_g \ge \bar F$ (Cor.~\ref{cor:beta_star}) \\
$\lambda$ & known lower bound on $P_{\mathrm{LCU}}$ fed to FPAA \\
$L$ & FPAA query complexity \\
$\delta$ & FPAA failure tolerance \\
\hline
\end{tabular}
\end{table}

\section{Imaginary-Time Evolution}
\label{app:ITE}

Formally replacing real time by imaginary time, $t = -i\beta$, in the Schr\"odinger equation $i\partial_t|\psi\rangle = \hat H|\psi\rangle$ yields the non-unitary imaginary-time equation $\partial_\beta|\psi(\beta)\rangle = -\hat H|\psi(\beta)\rangle$. The unnormalized solution is $e^{-\beta\hat H}|\psi_0\rangle$~\cite{goldberg1967integration, bader2013solving}. This is a formal analytic continuation, not unitary physical dynamics; the real parameter $\beta$ introduced here is referred to as imaginary time. For clarity, the derivation below is first written for a non-degenerate ground state with $p_0 > 0$. The degenerate case is treated explicitly at the end of the appendix and handled throughout the main text via $\gamma_0 = \sum_{k \in \mathcal G}|c_k|^2$ and $\Pi_{\mathcal G}|\psi_0\rangle/\sqrt{\gamma_0}$ (cf.\ Proposition~\ref{thm:exact_identity} and Appendix~\ref{app:state_error}).

Theoretically, ITE can always find the ground state if its initial state has any non-zero overlap with the ground subspace. The proof is as follows.

For a $d$-dimensional time-independent Hamiltonian $\hat{H}$, let the eigenvalues and the corresponding eigenstates be expressed as $E_0 \leq E_1 \leq \cdots \leq E_{d-1}$ and $|E_0\rangle, \cdots, |E_{d-1}\rangle$. Using eigendecomposition, we can write the initial state as $|\psi_0\rangle = \sum_{k=0}^{d-1}\sqrt{p_k}e^{i\phi_{k}}|E_k\rangle$, where $\langle E_k|\psi_0\rangle$ is expressed as the amplitude $\sqrt{p_k}$ and the phase $e^{i\phi_k}$. Here, $p_k \in [0, 1]$ and $\sum_{k=0}^{d-1} p_k = 1$.

The output state of ITE is given by:
\begin{align}
\begin{split}
\label{eq:ITE_op}
    |\psi (\beta)\rangle = \frac{e^{-\beta \hat{H}}|\psi_0\rangle}{\sqrt{\langle\psi_0|e^{-2\beta \hat{H}}|\psi_0\rangle}}.
\end{split}
\end{align}

The evolved state is given as $e^{-\beta \hat{H}}|\psi_0\rangle = \sum_{k=0}^{d-1} \sqrt{p_k} e^{-\beta E_k} e^{i\phi_{k}} |E_k\rangle$, with the normalization constant $\langle \psi_0|e^{-2\beta \hat{H}}|\psi_0\rangle = \sum_{k=0}^{d-1} p_k e^{-2\beta E_k}$.

By substituting the evolved state and normalization constant into Eq.~\eqref{eq:ITE_op}, we obtain:
\begin{align}
\begin{split}
     |\psi (\beta)\rangle
    = & \frac{\sum_{k=0}^{d-1}\sqrt{p_k}e^{-\beta E_k}e^{i\phi_{k}}|E_k\rangle}{\sqrt{\sum_{k=0}^{d-1}p_ke^{-2\beta E_k}}}\\
    = & \frac{e^{i\phi_0}|E_0\rangle + \sum_{k=1}^{d-1}\sqrt{\frac{p_k}{p_0}}e^{-\beta (E_k-E_0)}e^{i\phi_k }|E_k\rangle }{\sqrt{1 + \sum_{k=1}^{d-1}\frac{p_k}{p_0}e^{-2 \beta (E_k-E_0)}}}.
\end{split}
\end{align}
Here, we assume $p_0 > 0$ and a non-degenerate ground state.

For every excited eigenstate in the non-degenerate case, i.e.\ for every $k \ge 1$ with $E_k > E_0$, we have $e^{-\beta(E_k - E_0)} \to 0$ as $\beta \to \infty$. Consequently, $\lim_{\beta \to \infty} |\psi(\beta)\rangle = e^{i\phi_0}|E_0\rangle$.

If the ground subspace is degenerate, let $\mathcal G = \{k : E_k = E_0\}$, $\Pi_{\mathcal G} = \sum_{k \in \mathcal G}|E_k\rangle\langle E_k|$, and $\gamma_0 = \|\Pi_{\mathcal G}|\psi_0\rangle\|^2$. When $\gamma_0 > 0$, the same algebra gives
\begin{align}
\lim_{\beta \to \infty} \frac{e^{-\beta \hat H}|\psi_0\rangle}{\|e^{-\beta \hat H}|\psi_0\rangle\|} \;=\; \frac{\Pi_{\mathcal G}|\psi_0\rangle}{\sqrt{\gamma_0}}.
\end{align}
Thus excited-state components decay exponentially, while the relative amplitudes and phases among the ground-subspace components are preserved. The $\beta \to \infty$ limit is the normalized ground-subspace projection of the initial state, not a canonical choice of ground eigenstate. When $\gamma_0 = 0$, imaginary-time evolution cannot create ground-subspace support; the normalized state instead converges to the lowest-energy eigenspace having nonzero initial overlap.

This demonstrates that whenever the initial state has nonzero overlap with the ground subspace ($\gamma_0 > 0$, reducing to $p_0 > 0$ in the non-degenerate case), normalized ITE converges to the ground subspace in the limit of large $\beta$.

\section{Linear Combination of Unitaries}
\label{app:LCU}

Childs and Wiebe~\cite{childs2012Hamiltonian} demonstrated the Linear Combination of Unitaries (LCU) technique on a quantum computer. For two unitary matrices $U_a, U_b \in \mathbb{C}^{2^n \times 2^n}$, LCU probabilistically implements a normalized combination proportional to $\kappa U_a + U_b$ via ancilla post-selection, where $\kappa \geq 0$ is an arbitrary non-negative coefficient.

\begin{figure}[!hbp]
	\centering
	\includegraphics[width=7cm]{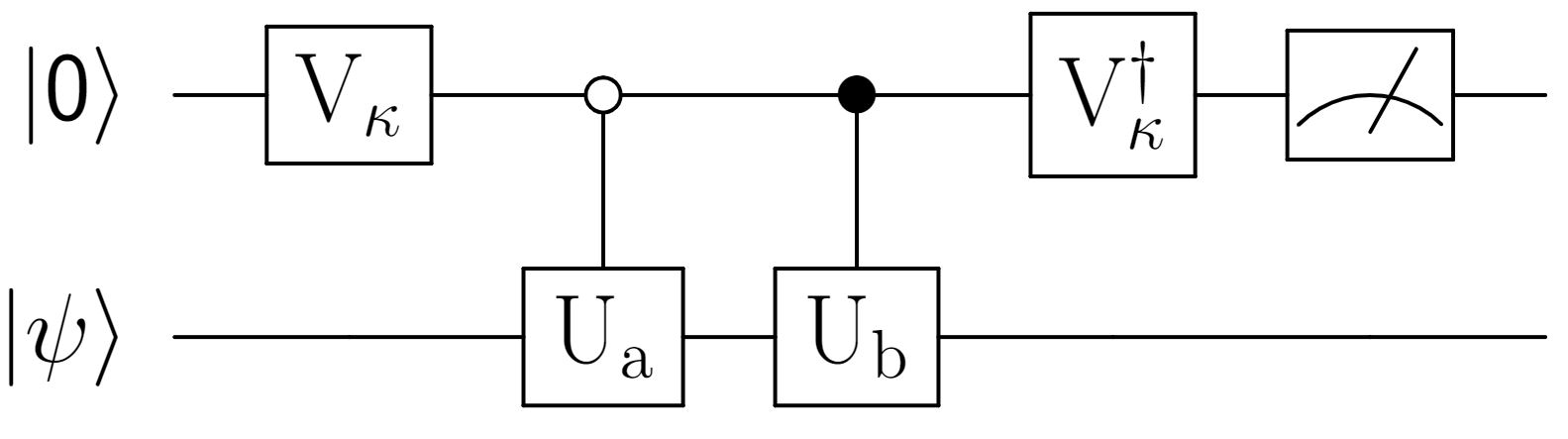}
	\caption{Quantum circuit for performing LCU by a non-deterministic process. We obtain $\kappa U_a + U_b$ up to normalization if the measurement result of the ancilla qubit is $|0\rangle$. Adapted from Ref.~\cite{childs2012Hamiltonian}.}
	\label{fig:lcu_circuit}
\end{figure}

If we define $V_{\kappa}$ as:
\begin{align}
\begin{split}
    V_{\kappa} \equiv
    \begin{pmatrix}
    \sqrt{\frac{\kappa}{\kappa + 1}} & \frac{-1}{\sqrt{\kappa + 1}}\\
    \frac{1}{\sqrt{\kappa + 1}} & \sqrt{\frac{\kappa}{\kappa + 1}}
    \end{pmatrix},
\end{split}
\end{align}
the complete process for LCU is shown in Fig.~\ref{fig:lcu_circuit}.

The entire process involves the following transformations:
\begin{align}
\begin{split}
    |0\rangle |\psi\rangle \mapsto
    & \left(\sqrt{\frac{\kappa}{\kappa + 1}}|0\rangle + \frac{1}{\sqrt{\kappa + 1}}|1\rangle\right)|\psi\rangle \\
    \mapsto&  \left(\sqrt{\frac{\kappa}{\kappa + 1}}|0\rangle U_a|\psi\rangle + \frac{1}{\sqrt{\kappa + 1}}|1\rangle U_b|\psi\rangle\right) \\
    \mapsto& |0\rangle \left(\frac{\kappa}{\kappa + 1}U_a + \frac{1}{\kappa + 1}U_b\right)|\psi\rangle + |1\rangle \frac{\sqrt{\kappa}}{\kappa + 1}(U_b - U_a)|\psi\rangle.
\end{split}
\end{align}

As a result, $\left(\frac{\kappa}{\kappa + 1}U_a + \frac{1}{\kappa + 1}U_b\right)$ is successfully executed when the ancilla qubit is measured as $|0\rangle$. If the ancilla qubit is measured as $|1\rangle$, the LCU fails with a probability $P_f$ of:
\begin{align}
\begin{split}
    P_{f}\leq \frac{\|U_a - U_b\|_2^2\,\kappa}{(\kappa + 1)^2}.
\end{split}
\end{align}
Specializing $U_a = I$, $U_b = -\mathrm{sgn}(x_{\boldsymbol\mu})\,\sigma_{\boldsymbol\mu}$, and $\kappa_{\boldsymbol\mu} = \coth(|\beta x_{\boldsymbol\mu}|) \ge 0$ gives the term-by-term worst-case bound
\begin{align}
P_{\mathrm{success}}(\beta x_{\boldsymbol\mu}) \;\ge\; 1 - \frac{4\kappa_{\boldsymbol\mu}}{(\kappa_{\boldsymbol\mu}+1)^2} \;=\; e^{-4\beta |x_{\boldsymbol\mu}|},
\end{align}
where the closed form follows from direct substitution of $\kappa_{\boldsymbol\mu} = \coth(\beta|x_{\boldsymbol\mu}|)$. Multiplying the term-by-term lower bounds over all nonzero-coefficient terms yields the corresponding worst-case joint lower bound
\begin{align}
P_{\mathrm{LCU}}(\beta) \;\ge\; \prod_{\boldsymbol\mu} e^{-4\beta |x_{\boldsymbol\mu}|} \;=\; e^{-4\beta W}.
\end{align}

\emph{Remark.} The term-by-term bound above is a state-independent worst-case expression. The main text replaces it with the exact state-dependent identity $P_{\mathrm{LCU}}(\beta) = e^{-2\beta W}\sum_k |c_k|^2 e^{-2\beta E_k}$ of Proposition~\ref{thm:exact_identity} (Sec.~\ref{FinITE}). The worst-case form is retained here because FPAA requires a lower bound on $P_{\mathrm{LCU}}$.

\section{Fixed-Point Amplitude Amplification}
\label{app:FPAA}

This appendix specifies the gate-level FPAA circuit together with the explicit phase schedule that achieves the query bound $L = \mathcal O(\log(2/\delta)/\sqrt{\lambda})$ quoted in the main text. Motivation and the generic success-probability bound $P_L \ge 1-\delta^2$ are stated in the FPAA primer (Sec.~\ref{subsec:FPAA_primer}); the FinITE specialization, with $\lambda$ any admissible lower bound on $P_{\mathrm{LCU}}(\beta)$, and the resulting query complexity are developed in Sec.~\ref{FinITE} and Appendix~\ref{app:cnot_complexity}.

\begin{figure}[!htbp]
	\centering
	\includegraphics[width=14cm]{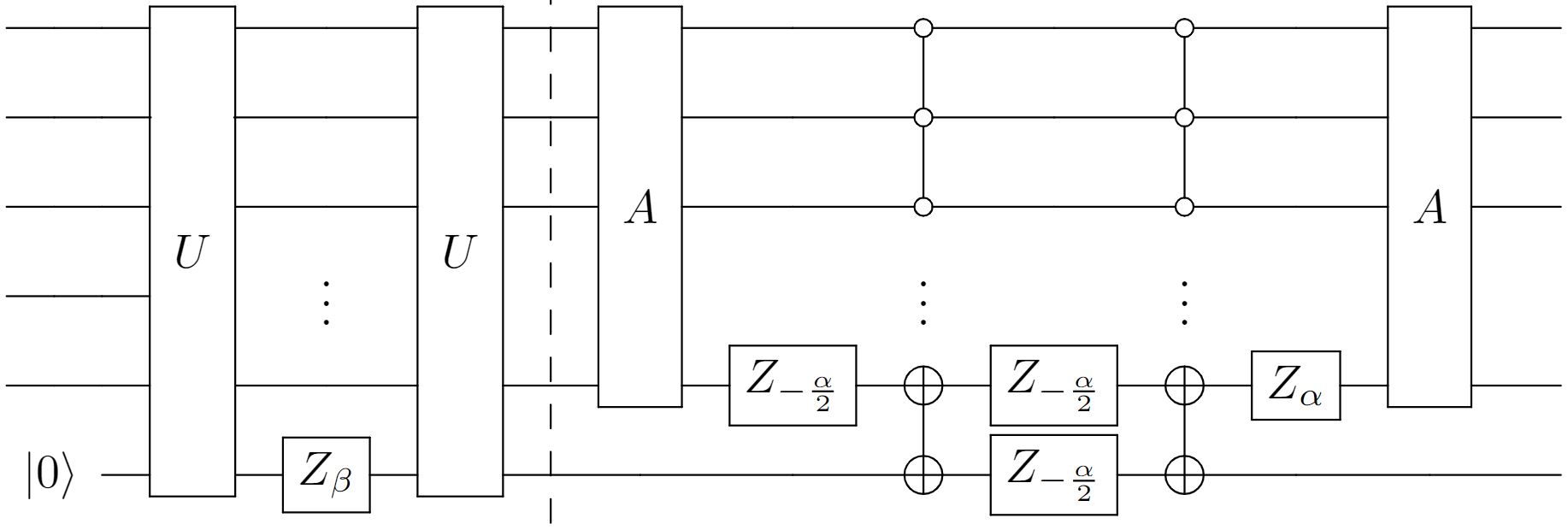}
	\caption{Quantum circuit for fixed-point amplitude amplification. $A$ prepares the initial state $|s\rangle = A|0\rangle^{\otimes n}$; $U$ is the good-subspace phase oracle (realized via phase kickback from a bit-flip ancilla oracle, with the ancilla prepared in $|-\rangle$); $Z_\theta$ is a $z$-axis rotation. Adapted from Ref.~\cite{yoder2014fixed}.}
	\label{fig:FPQAA_circuit}
\end{figure}

The circuit (Fig.~\ref{fig:FPQAA_circuit}) is built from three elementary pieces: the initial-state preparation $A$ with $|s\rangle = A|0\rangle^{\otimes n}$; a phase oracle $U$ that marks the good subspace by $U|g\rangle = -|g\rangle$ and $U|b\rangle = |b\rangle$ for $\langle g|b\rangle = 0$; and a single-qubit $z$-rotation $Z_\theta$. When the oracle is available only as a bit-flip on an ancilla, preparing the ancilla in $|-\rangle = (|0\rangle - |1\rangle)/\sqrt{2}$ converts the bit-flip into the required phase-flip via phase kickback, so the two oracle conventions yield identical FPAA output.

The FPAA phase schedule consists of two length-$l$ sequences $\{\alpha_j\}$ and $\{\beta_j\}$ applied to the generalized reflections about $|s\rangle$ and about the good subspace, respectively, across $l$ iterations of the generalized Grover iterate ($L = 2l+1$). Introducing $\gamma = 1/T_{1/L}(1/\delta)$, set by the target failure tolerance $\delta$ and the query depth $L$, with $T_L(x) = \cos(L\arccos x)$ the $L$-th Chebyshev polynomial of the first kind, the two sequences are phase-matched, $\beta_{l-j+1} = -\alpha_j$, with
\begin{equation}
\alpha_j \;=\; 2\,\operatorname{arccot}\!\left(\tan\!\frac{2\pi j}{L}\,\sqrt{1 - \gamma^2}\right), \qquad j = 1,\dots,l.
\label{eq:fpaa_phases}
\end{equation}
This schedule yields the success probability
\begin{equation}
P_L \;=\; 1 - \delta^2\,\Bigl[T_L\!\left(\sqrt{1 - \lambda}/\gamma\right)\Bigr]^2,
\label{eq:fpaa_pL}
\end{equation}
which attains $P_L \ge 1 - \delta^2$ on the admissible range $\lambda \ge 1 - \gamma^2$. Inverting this range in the $L \to \infty$, $\delta \to 0$ limit via $1 - \gamma^2 \sim (\log(2/\delta)/L)^2$ reproduces the query bound $L = \mathcal O(\log(2/\delta)/\sqrt{\lambda})$~\cite{yoder2014fixed}.

\section{Exact Fidelity--Success Identity: Derivation}
\label{app:exact_identity}

We prove Proposition~\ref{thm:exact_identity} and Corollaries~\ref{cor:gap_bound}--\ref{cor:beta_star}. The argument uses only the commutativity of the Pauli strings that compose $\hat H$ and the termwise-LCU normalization established in Sec.~\ref{imp_qc}.

\subsection{Block-encoded operator $\hat M(\beta)$ and its norm bound}
Fix a Pauli decomposition $\hat H = \sum_{\boldsymbol\mu} x_{\boldsymbol\mu} \sigma_{\boldsymbol\mu}$ over an $n$-qubit Hilbert space with real coefficients $x_{\boldsymbol\mu} \in \mathbb R$ and pairwise-commuting Pauli strings $[\sigma_{\boldsymbol\mu}, \sigma_{\boldsymbol\nu}] = 0$. Following Sec.~\ref{sec:background} and the conventions of Proposition~\ref{thm:exact_identity}, the identity term has been removed and absorbed into the classical scalar $e^{-\beta c}$, so $\hat H$ here denotes the identity-removed part. We restrict the index set $\boldsymbol\mu$ to nonzero-coefficient terms, so $\mathrm{sgn}(x_{\boldsymbol\mu})$ is well defined, and take $\beta \ge 0$. The commuting Hermitian set $\{\sigma_{\boldsymbol\mu}\}$ admits an orthonormal common eigenbasis $\{|E_k\rangle\}$. Since $\hat H$ is their real linear combination, $\{|E_k\rangle\}$ is also an eigenbasis of $\hat H$. Write the normalized initial state as $|\psi_0\rangle = \sum_k c_k |E_k\rangle$ with $\sum_k |c_k|^2 = 1$. The exact termwise-LCU block-encodes the scaled operator
\begin{align}
\hat M(\beta) \; := \; m(\beta)\, e^{-\beta \hat H}, \qquad m(\beta) \; := \; \prod_{\boldsymbol\mu \ne 0} e^{-|\beta x_{\boldsymbol\mu}|} \;=\; e^{-\beta W}, \qquad W \; := \; \sum_{\boldsymbol\mu \ne 0} |x_{\boldsymbol\mu}|,
\end{align}
which is precisely the scaling factor appearing in Eq.~\eqref{exact_trotterized_more_expand} after absorbing the $\exp(|\beta x_{\boldsymbol\mu}|)$ denominators of each term. Each exact single-term block-encodes the contraction $e^{-|\beta x_{\boldsymbol\mu}|}\,e^{-\beta x_{\boldsymbol\mu}\sigma_{\boldsymbol\mu}} = \alpha_{\boldsymbol\mu} I - \gamma_{\boldsymbol\mu}\,\mathrm{sgn}(x_{\boldsymbol\mu})\,\sigma_{\boldsymbol\mu}$, where $\alpha_{\boldsymbol\mu} = (1 + e^{-2\beta |x_{\boldsymbol\mu}|})/2 \in [1/2, 1]$ and $\gamma_{\boldsymbol\mu} = (1 - e^{-2\beta |x_{\boldsymbol\mu}|})/2 \in [0, 1/2]$. The weights $\alpha_{\boldsymbol\mu}$ and $\gamma_{\boldsymbol\mu}$ are nonnegative and sum to one, so each block is a convex combination of the two unitaries $I$ and $-\mathrm{sgn}(x_{\boldsymbol\mu})\sigma_{\boldsymbol\mu}$ and therefore has operator norm at most $1$. By submultiplicativity of the operator norm,
\begin{align}
\Bigl\|\prod_{\boldsymbol\mu} \bigl(e^{-|\beta x_{\boldsymbol\mu}|}\,e^{-\beta x_{\boldsymbol\mu}\sigma_{\boldsymbol\mu}}\bigr)\Bigr\|_2 \;\le\; \prod_{\boldsymbol\mu} \bigl\|e^{-|\beta x_{\boldsymbol\mu}|}\,e^{-\beta x_{\boldsymbol\mu}\sigma_{\boldsymbol\mu}}\bigr\|_2 \;\le\; 1;
\end{align}
Pairwise commutativity of the $\sigma_{\boldsymbol\mu}$ is then used only to identify this ordered block product with $m(\beta)\,e^{-\beta\hat H} = \hat M(\beta)$, so $\|\hat M(\beta)\|_2 \le 1$. Under the exact block-encoding/post-selection convention (ancilla initialized in $|0\rangle$, post-selected on $|0\rangle$), successful post-selection of every single-term ancilla applies $\hat M(\beta)$ coherently to $|\psi_0\rangle$.

\subsection{Proof of Eq.~\eqref{eq:P_LCU}}
By the LCU block-encoding convention of Sec.~\ref{imp_qc} (ancilla initialized in $|0\rangle$ and post-selected on $|0\rangle$), the LCU success probability equals the squared norm of the unnormalized post-selected state,
\begin{align}
P_{\mathrm{LCU}}(\beta) \;=\; \| \hat M(\beta) |\psi_0\rangle \|_2^2 \;=\; m(\beta)^2\, \langle \psi_0 | e^{-2\beta \hat H} | \psi_0 \rangle,
\end{align}
where Hermiticity of $\hat H$ gives $(e^{-\beta \hat H})^\dagger = e^{-\beta \hat H}$ so that $\hat M(\beta)^\dagger \hat M(\beta) = m(\beta)^2\, e^{-2\beta \hat H}$. Expanding in the orthonormal eigenbasis $\{|E_k\rangle\}$ and using $e^{-2\beta \hat H}|E_k\rangle = e^{-2\beta E_k}|E_k\rangle$,
\begin{align}
\langle \psi_0 | e^{-2\beta \hat H} | \psi_0 \rangle \;=\; \sum_k |c_k|^2\, e^{-2\beta E_k},
\end{align}
and substituting $m(\beta)^2 = e^{-2\beta W}$ yields Eq.~\eqref{eq:P_LCU}.

\subsection{Proof of Eq.~\eqref{eq:F_g}}
The normalized post-selected state is
\begin{align}
|\psi(\beta)\rangle \;=\; \frac{\hat M(\beta) |\psi_0\rangle}{\sqrt{P_{\mathrm{LCU}}(\beta)}} \;=\; \frac{1}{\sqrt{\sum_k |c_k|^2 e^{-2\beta E_k}}} \sum_k c_k\, e^{-\beta E_k} |E_k\rangle,
\end{align}
where the $m(\beta)$ factor cancels in the ratio. Let $\Pi_{\mathcal G} = \sum_{k \in \mathcal G} |E_k\rangle\langle E_k|$ denote the orthogonal projector onto the ground subspace $\mathcal G = \{k : E_k = E_0\}$. This projector is basis-invariant within $\mathcal G$, so the ground-subspace fidelity $F_g(\beta) = \langle \psi(\beta)|\Pi_{\mathcal G}|\psi(\beta)\rangle$ does not depend on the choice of orthonormal basis inside a degenerate $\mathcal G$. Evaluating the projector expectation,
\begin{align}
F_g(\beta) \;=\; \sum_{k \in \mathcal G} |\langle E_k | \psi(\beta)\rangle|^2 \;=\; \frac{\sum_{k \in \mathcal G} |c_k|^2\, e^{-2\beta E_k}}{\sum_j |c_j|^2\, e^{-2\beta E_j}} \;=\; \frac{\gamma_0\, e^{-2\beta E_0}}{\sum_j |c_j|^2\, e^{-2\beta E_j}},
\end{align}
using $E_k = E_0$ for $k \in \mathcal G$ and $\gamma_0 = \sum_{k \in \mathcal G} |c_k|^2$. This is Eq.~\eqref{eq:F_g}. Note that the formula is well defined whether $\mathcal G$ is a singleton or a degenerate subspace.

\subsection{Proof of the product identity Eq.~\eqref{eq:PF_identity}}
Multiplying Eqs.~\eqref{eq:P_LCU} and \eqref{eq:F_g} cancels the state-dependent partition sum:
\begin{align}
P_{\mathrm{LCU}}(\beta)\, F_g(\beta) \;=\; e^{-2\beta W} \cdot \gamma_0\, e^{-2\beta E_0} \;=\; \gamma_0\, e^{-2\beta(W + E_0)}.
\end{align}
The product depends only on the decomposition-level quantity $W$, the ground-state energy $E_0$, and the initial-state overlap $\gamma_0$. The detailed non-ground-state spectral data---the $\{E_k, |c_k|^2\}$ for $k \notin \mathcal G$---cancels between $P_{\mathrm{LCU}}$ and $F_g$.

\subsection{Proof of Corollary~\ref{cor:gap_bound} (gap bound on $F_g$)}
Factor $e^{-2\beta E_0}$ out of the denominator of Eq.~\eqref{eq:F_g}:
\begin{align}
F_g(\beta) \;=\; \frac{\gamma_0}{\gamma_0 + \sum_{k \notin \mathcal G} |c_k|^2\, e^{-2\beta(E_k - E_0)}}.
\end{align}
For $k \notin \mathcal G$, $E_k - E_0 \ge \Delta$ by definition of the gap. Combined with $\beta \ge 0$, this gives $e^{-2\beta(E_k - E_0)} \le e^{-2\beta \Delta}$. Using normalization $\sum_{k \notin \mathcal G} |c_k|^2 = 1 - \gamma_0$,
\begin{align}
\sum_{k \notin \mathcal G} |c_k|^2\, e^{-2\beta(E_k - E_0)} \;\le\; (1 - \gamma_0)\, e^{-2\beta \Delta},
\end{align}
Substituting into the factored $F_g$ yields Eq.~\eqref{eq:gap_bound}. The closed-form equivalence to $[1 + ((1-\gamma_0)/\gamma_0)\,e^{-2\beta\Delta}]^{-1}$ is immediate algebra for $\gamma_0 > 0$. For $\gamma_0 = 0$, the right-hand side of Eq.~\eqref{eq:gap_bound} is identically zero, consistent with $F_g(\beta) \ge 0$.

\subsection{Proof of Corollary~\ref{cor:beta_star} and the lower bound on $P_{\mathrm{LCU}}(\beta^\star)$}
For $\gamma_0 > 0$ and $\Delta > 0$, Eq.~\eqref{eq:gap_bound} admits the equivalent closed form $F_g(\beta) \ge [1 + ((1-\gamma_0)/\gamma_0)\,e^{-2\beta\Delta}]^{-1}$. Requiring this right-hand side to be at least $\bar F \in (0,1)$ gives
\begin{align}
1 + \frac{1 - \gamma_0}{\gamma_0}\, e^{-2\beta \Delta} \;\le\; \frac{1}{\bar F},
\end{align}
i.e.\ $\frac{1 - \gamma_0}{\gamma_0}\, e^{-2\beta \Delta} \le \frac{1 - \bar F}{\bar F}$. Solving for $\beta$, the bound is informative only when $\bar F > \gamma_0$, in which case the logarithm is positive and
\begin{align}
\beta \;\ge\; \frac{1}{2\Delta}\, \log\!\left( \frac{\bar F\,(1 - \gamma_0)}{\gamma_0\,(1 - \bar F)} \right).
\end{align}
This is Eq.~\eqref{eq:beta_star}. Finally, the product identity~\eqref{eq:PF_identity} combined with $F_g(\beta) \le 1$ gives the lower bound
\begin{align}
P_{\mathrm{LCU}}(\beta^\star) \;=\; \frac{\gamma_0\, e^{-2\beta^\star(W + E_0)}}{F_g(\beta^\star)} \;\ge\; \gamma_0\, e^{-2\beta^\star(W + E_0)},
\end{align}
which is Eq.~\eqref{eq:P_at_beta_star}. The matching upper bound $P_{\mathrm{LCU}}(\beta^\star) \le (\gamma_0/\bar F)\, e^{-2\beta^\star(W+E_0)}$ follows from $F_g(\beta^\star) \ge \bar F$. The corollary assumes $0 < \gamma_0 < 1$, $\Delta > 0$, and $\bar F \in (\gamma_0, 1)$, so the logarithm defining $\beta^\star$ is finite. If $\gamma_0 = 1$, then $F_g(\beta) \equiv 1$ and one may take $\beta^\star = 0$. If $\gamma_0 = 0$, then imaginary-time filtering cannot create ground-space support, so no target $\bar F > 0$ is achievable, consistent with $P_{\mathrm{LCU}}(\beta) F_g(\beta) = 0$. \qed

\section{Complexity Analysis}
\label{app:complexity}

In this section, we analyze the computational complexity of the FinITE algorithm with respect to state error and CNOT complexity.

\subsection{State Error}
\label{app:state_error}

For clarity, the following derivation is written for the non-degenerate ground state. The general (possibly degenerate) case is covered by Proposition~\ref{thm:exact_identity} and Corollary~\ref{cor:gap_bound}. With $\gamma_0 = \sum_{k \in \mathcal G}|c_k|^2 > 0$ the ground-space weight and $\Delta = \min_{k \notin \mathcal G}(E_k - E_0)$ the gap above $\mathcal G$, the state-preparation error bound takes the same form as below, with $p_0$ replaced by $\gamma_0$ and the single eigenvector $|E_0\rangle$ replaced by the normalized projection $\Pi_{\mathcal G}|\psi_0\rangle/\sqrt{\gamma_0}$.

We show the state error in the FinITE algorithm with the limit $\lim_{\beta \to \infty} |\psi(\beta)\rangle = e^{i\phi_0}|E_0\rangle$. Due to practical limitations, it is difficult to assign $\beta$ to infinity, even when dealing with a Hamiltonian where all terms commute.

We implement the ITE operator using an appropriate finite $\beta$, producing $\ket{\psi (\beta)}$, which can be expressed as follows:
\begin{align}
\begin{split}
     \ket{\psi (\beta)} &= \frac{e^{-\beta \hat H}\ket{\psi_0}}{\sqrt{\bra{\psi_0}e^{-2\beta \hat H}\ket{\psi_0}}} \\
    &= \frac{\sum_{k=0}^{d-1}\sqrt{p_k}e^{-\beta E_k}e^{i\phi_{k}}|E_k\rangle}{\sqrt{\sum_{k=0}^{d-1}p_ke^{-2\beta E_k}}} \\
    &= \frac{e^{i\phi_0}|E_0\rangle + \sum_{k=1}^{d-1}\sqrt{\frac{p_k}{p_0}}e^{-\beta (E_k-E_0)}e^{i\phi_k }|E_k\rangle }{\sqrt{1 + \sum_{k=1}^{d-1}\frac{p_k}{p_0}e^{-2 \beta (E_k-E_0)}}}.
\end{split}
\end{align}

The Hamiltonian corresponding to the combinatorial optimization problems we address, such as QUBO and HUBO, consists of mutually commuting terms. As a result, there is no need to consider Trotter error.

The upper bound of the error $\|\ket{\psi (\beta)} - e^{i\phi_0}|E_0\rangle \|_2 = \sqrt{2 - 2\Re(\langle\psi (\beta)|e^{i\phi_0}|E_0\rangle)}$ is as follows:
\begin{align}
\begin{split}
\label{eq:state_error_cal}
     \|\ket{\psi (\beta)} - e^{i\phi_0}|E_0\rangle \|_2
     &= \sqrt{2 - 2 \frac{1}{\sqrt{1 + \sum_{k=1}^{d-1}\frac{p_k}{p_0}e^{-2 \beta (E_k-E_0)}}}}\\
     &\leq \sqrt{2 - 2 \frac{1}{\sqrt{1 + e^{-2 \beta (E_1-E_0)}\sum_{k=1}^{d-1}\frac{p_k}{p_0}}}}\\
     &= \sqrt{2 - 2 \frac{1}{\sqrt{1 + \frac{1-p_0}{p_0}e^{-2 \beta (E_1-E_0)}}}} \leq \epsilon .
\end{split}
\end{align}

Solving the inequality for $\beta$ gives
\begin{align}
\begin{split}
     \beta \;\ge\; \frac{1}{2\left(E_1-E_0\right)}\, \log\!\left[\, \frac{(1 - p_0)/p_0}{\left(1-\frac{\epsilon^2}{2}\right)^{-2} - 1} \,\right].
\end{split}
\end{align}
In the nontrivial regime $0 < p_0 < 1$ and $0 < \epsilon < \sqrt{2 - 2\sqrt{p_0}}$, the logarithm on the right is positive. For $\epsilon \ge \sqrt{2 - 2\sqrt{p_0}}$, $\beta = 0$ already satisfies the error target.

\subsection{CNOT Complexity}
\label{app:cnot_complexity}

We asymptotically calculate the CNOT complexity for our algorithm. The PUBO Hamiltonians considered here consist of pairwise-commuting Pauli-$Z$ terms, so no Trotter error arises. For convenience, we refer to the total number of Pauli string terms in the Hamiltonian as $M$.

The $M$ LCU blocks, one for each Pauli term, collectively implement $m(\beta)\,e^{-\beta\hat H}$. Each block uses a $k$-local controlled Pauli-string operation, contributing $\mathcal O(k)$ CNOTs, so the full LCU pass costs $\mathcal O(Mk)$ CNOTs.

We also use the fixed-point amplitude amplification method to amplify the probability of success. Applying the process of Fig.~\ref{fig:FPQAA_circuit} to LCU, we account for two auxiliary oracles: the good-state oracle $U$, which marks LCU-success states (all $M$ LCU ancillas in $|0\rangle$), and the initial-state reflection $S_0 = 2|0\rangle^{\otimes(M+n)}\langle 0|^{\otimes(M+n)} - I$, which reflects about the full all-zero initial state on the combined $M+n$ qubit register. Each is implemented by a multi-controlled NOT with phase kickback from an auxiliary $|-\rangle$ ancilla (Appendix~\ref{app:FPAA}).

\begin{figure}[!htbp]
	\centering
	\includegraphics[width=7cm]{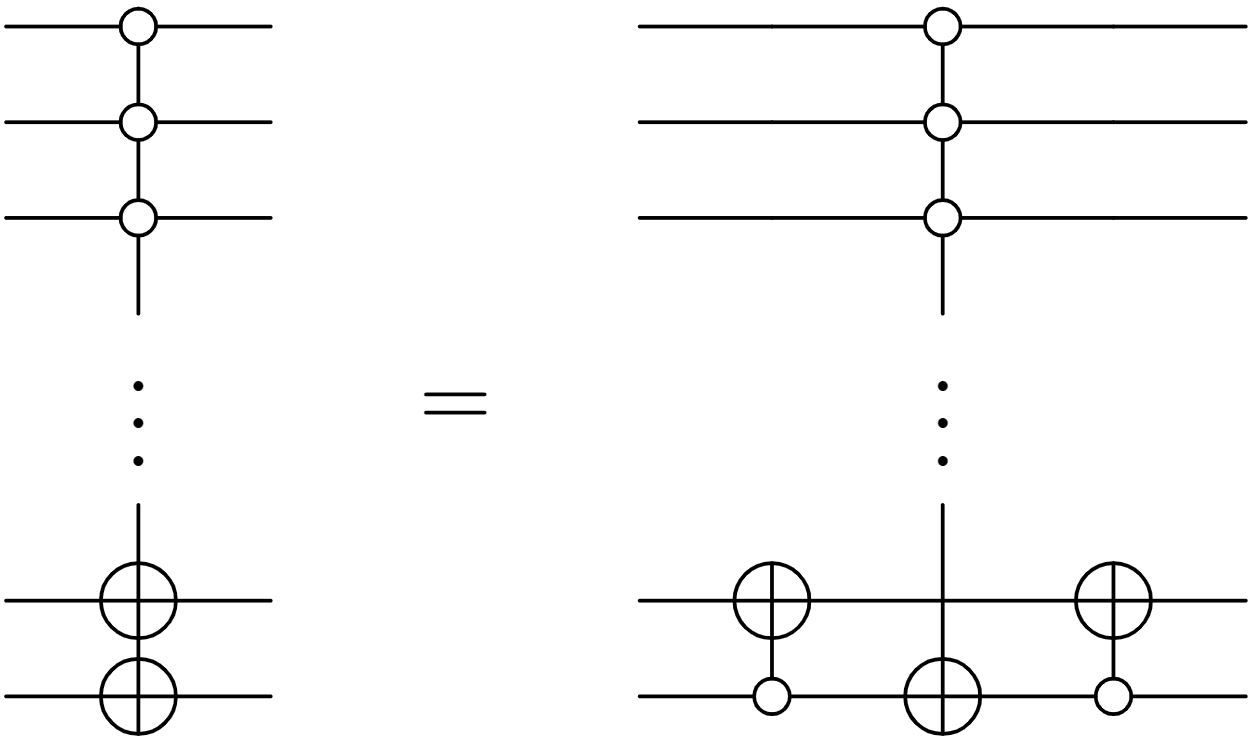}
	\caption{Quantum circuit for a multi-controlled double NOT gate. The gate decomposes into two CNOT gates and one multi-controlled NOT gate.}
	\label{fig:MCDN_gate}
\end{figure}

The good-state oracle $U$ uses a multi-controlled NOT gate with $M$ controls (one for each LCU ancilla), requiring $\mathcal O(M^2)$ CNOT gates in an ancilla-free decomposition~\cite{silva2022linear}. The reflection $S_0$ uses a multi-controlled NOT on $M+n$ controls, costing $\mathcal O((M+n)^2)$ CNOT gates. Fig.~\ref{fig:MCDN_gate} shows the two-target form used here, which adds two CNOT gates on top of the multi-controlled NOT. The Hamiltonian dimension is $2^n \times 2^n$.

One FPAA iteration applies $A$ and $A^\dagger$, together contributing $\mathcal O(Mk)$ CNOT gates. The good-state oracle $U$ contributes $\mathcal O(M^2)$, and the reflection $S_0$ contributes $\mathcal O((M+n)^2)$. Each iteration therefore costs $\mathcal O(Mk + M^2 + (M+n)^2)$ CNOTs to leading order. Over $L$ iterations together with the initial-state preparation $A$, the CNOT count is $\mathcal O(L\,[Mk + M^2 + (M+n)^2])$. This count assumes the initial-state preparation cost is subsumed in $A$ (e.g., $|+\rangle^{\otimes n}$ requires only $\mathcal O(n)$ single-qubit Hadamards); a nontrivial warm start with CNOT cost $C_{\mathrm{prep}}$ adds $\mathcal O(L\,C_{\mathrm{prep}})$ separately. The query complexity $L$ obeys the FPAA bound $L = \mathcal O(\log(2/\delta)/\sqrt{\lambda})$~\cite{yoder2014fixed}, where $\delta$ is the target FPAA failure tolerance and $\lambda$ is any lower bound on $P_{\mathrm{LCU}}(\beta)$. Combining with Eq.~\eqref{eq:P_at_beta_star}, $\lambda = \gamma_0\, e^{-2\beta^\star(W + E_0)}$ is admissible at the fidelity threshold $\beta^\star$, giving the $\beta$-resolved total
\begin{equation}
\#\,\mathrm{CNOT} \;=\; \mathcal O\!\left(\frac{\log(2/\delta)\, e^{\beta^\star(W + E_0)}}{\sqrt{\gamma_0}}\,\bigl[Mk + M^2 + (M+n)^2\bigr]\right).
\end{equation}
The exponential factor $e^{\beta^\star(W+E_0)}$ is the structural cost of ITE within this termwise-LCU implementation, identified by Proposition~\ref{thm:exact_identity}. Amplitude amplification reduces the repetition cost from $\mathcal O(1/P_{\mathrm{LCU}})$ to $\mathcal O(1/\sqrt{P_{\mathrm{LCU}}})$, but does not remove the $e^{\beta^\star(W+E_0)}$ factor inherent in the $1/\sqrt{P_{\mathrm{LCU}}}$ dependence.

The total qubit count is $n + M + 1$: $n$ system qubits, $M$ LCU ancillas (one for each Pauli term), and one phase-kickback scratch ancilla used for the multi-controlled reflections. Including one- and two-qubit gate counts together does not change the asymptotic scaling: the $V_\kappa$ single-qubit rotations contribute $\mathcal O(M)$ additional single-qubit gates for each LCU pass, subsumed by the overall $\mathcal O(Mk + M^2 + (M+n)^2)$ scaling.
\end{document}